\documentclass{elsart}
\usepackage{graphicx}
\usepackage{amssymb}
\begin{document}

\begin{frontmatter}

\title{Search for Global Dipole Enhancements in the HiRes-I Monocular Data 
above ${\bf 10^{18.5}}$~eV} 

\author[utah]{R.~Abbasi}
\author[utah]{T.~Abu-Zayyad,}
\author[lanl]{J.F.~Amann,}
\author[utah]{G.~Archbold,}
\author[adelaide]{J.A.~Bellido,}
\author[utah]{K.~Belov,}
\author[umt]{J.W.~Belz,}
\author[rutgers]{D.R.~Bergman,}
\author[utah]{Z.~Cao,}
\author[adelaide]{R.W.~Clay,}
\author[nevis]{B.~Connolly,}
\author[lanl]{M.D.~Cooper,}
\author[adelaide]{B.R.~Dawson,}
\author[nevis]{C.~Finley,}
\author[utah]{W.F.~Hanlon,}
\author[lanl]{C.M.~Hoffman,}
\author[lanl]{M.H.~Holzscheiter,}
\author[utah]{P.~H\"{u}ntemeyer,}
\author[utah]{C.C.H.~Jui,}
\author[utah]{K.~Kim,}
\author[umt]{M.A.~Kirn,}
\author[utah]{E.C.~Loh,}
\author[icrr]{N.~Manago,}
\author[lanl]{L.J.~Marek,}
\author[utah]{K.~Martens,}
\author[unm]{G. Martin,}
\author[unm]{J.A.J.~Matthews,}
\author[utah]{J.N.~Matthews,}
\author[lanl]{C.A.~Painter,}
\author[rutgers]{L.~Perera,}
\author[utah]{K.~Reil,}
\author[utah]{R.~Riehle,}
\author[unm]{M.~Roberts,}
\author[lanl]{J.S.~Sarracino,}
\author[icrr]{M.~Sasaki,}
\author[rutgers]{S.R.~Schnetzer,}
\author[adelaide]{K.M.~Simpson,}
\author[lanl]{C.~Sinnis,}
\author[utah]{J.D.~Smith,}
\author[utah]{P.~Sokolsky,}
\author[nevis]{C.~Song,}
\author[utah]{R.W.~Springer,}
\author[utah]{B.T.~Stokes\corauthref{cor1},}
\author[utah]{S.B.~Thomas,}
\author[lanl]{T.N.~Thompson,}
\author[rutgers]{G.B.~Thomson,}
\author[lanl]{D.~Tupa,}
\author[nevis]{S.~Westerhoff,}
\author[utah]{L.R.~Wiencke,}
\author[rutgers]{A.~Zech,}
\author[nevis]{and X.~Zhang}
\collaboration{The High Resolution Fly's Eye Collaboration}
\address[utah]{University of Utah,
Department of Physics and High Energy Astrophysics Institute,
Salt Lake City, Utah, USA}
\address[lanl]{Los Alamos National Laboratory,
Los Alamos, NM, USA}
\address[adelaide]{University of Adelaide, Department of Physics,
Adelaide, South Australia,  Australia}
\address[umt]{University of Montana, Department of Physics and Astronomy,
Missoula, Montana, USA.}
\address[rutgers]{Rutgers --- The State University of New Jersey,
Department of Physics and Astronomy,
Piscataway, New Jersey, USA}
\address[nevis]{Columbia University, Department of Physics and
Nevis Laboratory, New York, New York, USA}
\address[icrr]{University of Tokyo,
Institute for Cosmic Ray Research,
Kashiwa, Japan}
\address[unm]{University of New Mexico,
Department of Physics and Astronomy,
Albuquerque, New Mexico, USA  }
\corauth[cor1]{
Corresponding~author.\ {\it E-mail~address}:~stokes@cosmic.utah.edu~(B.T.~Stokes)
}

\newpage

\begin{abstract}
Several proposed source models for Ultra-High Energy Cosmic Rays (UHECRs) 
consist of dipole distributions oriented towards major astrophysical landmarks
such as the galactic center, M87, or Centaurus A.  We use 
a comparison between real data and simulated data to show that the
HiRes-I monocular data for energies above  $10^{18.5}$~eV is, in
fact, consistent with an {\it isotropic} source model.  We then explore 
methods to quantify our sensitivity to dipole source models oriented towards
the Galactic Center, M87, and Centaurus A. 
\end{abstract}

\end{frontmatter}

\section{Introduction}

The observation of Ultra-High Energy Cosmic Rays (UHECRs) has now spanned 
over forty years.  Over that period, many source models have been 
proposed to explain the origin of these remarkable events. In the past five 
years, theoretical models have been suggested that would potentially produce 
dipole distributions oriented towards M87 \cite{Biermann:fd} or
Centaurus A \cite{Farrar:2000nw,Anchordoqui:2001nt}.  In addition, the 
Akeno Giant Air Shower Array (AGASA) has reported findings
suggesting a 4\% dipole-like enhancement oriented towards the Galactic Center 
present in its events with energies around 
$10^{18}$~eV \cite{Hayashida:1999ab}.  This result seemed to be
corroborated by findings published by the Fly's Eye experiment in 1999 that  
suggested the possibility of an 
enhancement in the galactic plane also at energies around $10^{18}$~eV
 \cite{Bird:1998nu}, and also
by a re-analysis of data from the SUGAR array that was published in 2001 
\cite{Bellido:2000tr} that showed an enhancement in the general vicinity of
the Galactic Center.  

However, both AGASA and Fly's Eye are subject to a limiting factor; 
they are both located too far north in latitude to directly 
observe the Galactic Center 
itself.  The re-analysis of SUGAR data actually demonstrated an excess 
that was offset from the Galactic Center by $7.5^\circ$ and was more consistent
with a point source than a global dipole effect \cite{Bellido:2000tr}.
While the current High Resolution Fly's Eye (HiRes) experiment 
is subject to a similar 
limitation in sky coverage as the AGASA and Fly's Eye experiments, 
we will show that, by properly estimating the HiRes aperture and
angular resolution, we can effectively exclude these dipole source  
models to a certain degree of sensitivity.  However, we are not able to 
completely exclude the findings of AGASA or the theoretical predictions
mentioned above.

Our methods for detecting the presence of a dipole source model will
be based upon comparisons between the real data and a large quantity of 
events generated by our Monte Carlo
simulation program.  The simulated data possess
the same aperture and exposure as the actual HiRes-I monocular data set.
In order to measure the presence of a dipole effect in our event
sample, we use first a conventional binning technique that considers the event 
counts for the full range of opening angles from the center of each 
proposed dipole distribution.  
We then show how the asymmetric angular resolution of a monocular air
fluorescence detector can be accommodated in this method.
We ascertain the 90\% confidence interval for a dipole source model for
each of the three dipoles considered by comparing our real data with large 
numbers of similar-sized simulated data sets.  We then consider 
the effects of systematic uncertainties on our measurements.
To conclude, we use a novel technique measuring the  
information dimension \cite{stokes}, $D_{\rm I}$, of our sample to place 
an independent 90\% confidence interval on the scaling parameter, $\alpha$,
that we use to quantify our dipole source model. 

\section{The Dipole Function}

A dipole source model can be described, as first proposed by Farrar and Piran
\cite{Farrar:2000nw}:
\begin{equation}
n=\frac{1}{2}+\frac{\alpha}{2}\cos\theta,
\label{equation:dipole}
\end{equation}
where $n$ is the relative density of cosmic rays in a given direction, 
$\theta$ is the opening angle between that direction and the global 
maximum of the distribution, and $\alpha$ is the customary anisotropy amplitude
\cite{Sommers:2000us}:
\begin{equation}
\alpha=\frac{n_{\rm max}-n_{\rm min}}{n_{\rm max}+n_{\rm min}}.
\end{equation}
The cases of 
$\alpha=1$ and $\alpha=-1$ correspond to 
100\% dipole distributions in the direction of the center and anti-center
of the dipole source model, respectively.  
The case of $\alpha=0$ corresponds to an isotropic source model.

A simple scheme for measuring $\alpha$ consists of constructing a 
{\it dipole function} in the following manner:
\begin{enumerate}
\item The opening angle is measured between the arrival direction
of an event and the center of the proposed dipole source model.
\item The cosine of the opening angle is then histogrammed.
\item The preceding steps are repeated until all of the events are considered.
\item The resulting curve produced by the histogram is the dipole function.
\end{enumerate}
The dipole function has two variable parameters:  the bin width, 
$\Delta(\cos\theta)$, and the total number of counts in all of the bins.  At
first glance, it would seem that the total bin count is fixed upon the 
total number of events, but we will show that this isn't necessarily the case
when we consider how to accommodate angular resolution. 

In the simplest case of a sample 
that contains a very large number of events with a 
constant exposure and aperture over the entire sky, the dipole function will
be proportional to equation \ref{equation:dipole}.  
We propose two simple ways that one can
quantify the dipole function for this sample; the most obvious way is to 
consider its slope.  We can see by referring to equation \ref{equation:dipole}
 that this is equal to $\frac{\alpha}{2}$.  A second way of quantifying 
$\alpha$ is to consider the mean cosine value, $<\!\!\cos\theta\!\!>$ 
for the dipole function:
\begin{equation}
<\!\!\cos\theta\!\!>=\frac{1}{2}\int_{-1}^{1}\cos\theta(1+\alpha\cos\theta)\, 
d(\cos\theta)=\frac{1}{3}\alpha.
\label{equation:mean}
\end{equation}
Both methods of quantification produce 
values that are dependent upon $\alpha$.  While the dependence of $<\!\!\cos\theta\!\!>$
is linear in $\alpha$ for the case of homogeneous full-sky coverage, we
will find that this is not necessarily the case when considering the cumulative
exposure of a ground-based air fluorescence detector.

\section{Calculating the Dipole Function for the HiRes-I Monocular Data}

We now consider the real data sample consisting of events
that were included in the HiRes-I monocular spectrum measurement 
\cite{Abu-Zayyad:2002ta,Abu-Zayyad:2002sf}.
This set contains 1526 events observed between May~1997 and 
February~2003 with measured energies greater than $10^{18.5}$~eV.  
The HiRes monocular data set represents a cumulative 
exposure of $\sim3000$~km$^2\cdot$sr$\cdot$yr at $5\times10^{19}$~eV.
\begin{figure}[t,b,p]
\begin{center}
\begin{tabular}{c@{\hspace{0.0cm}}c}
(a)\includegraphics[width=6.15cm]{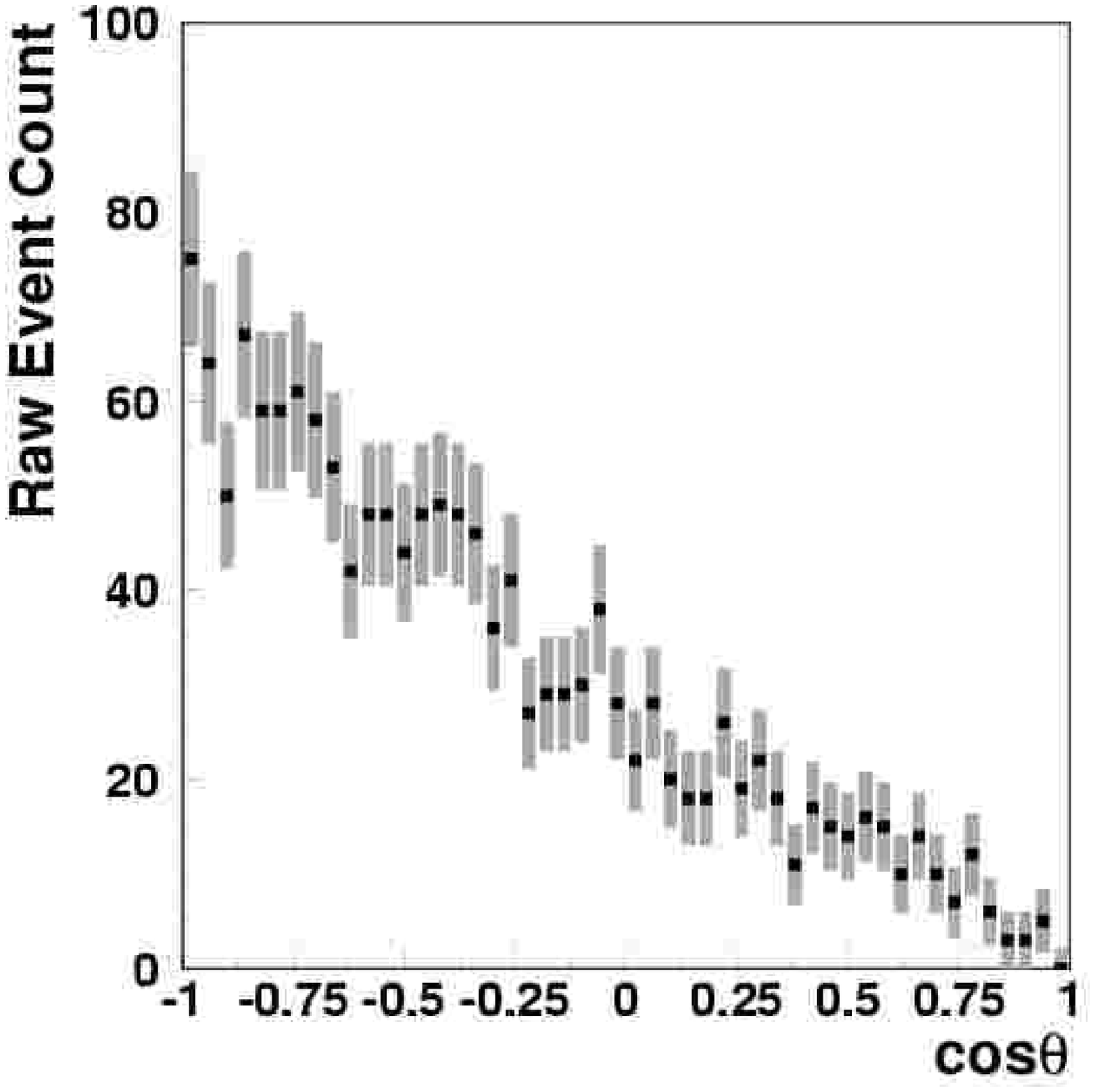}&
(b)\includegraphics[width=6.15cm]{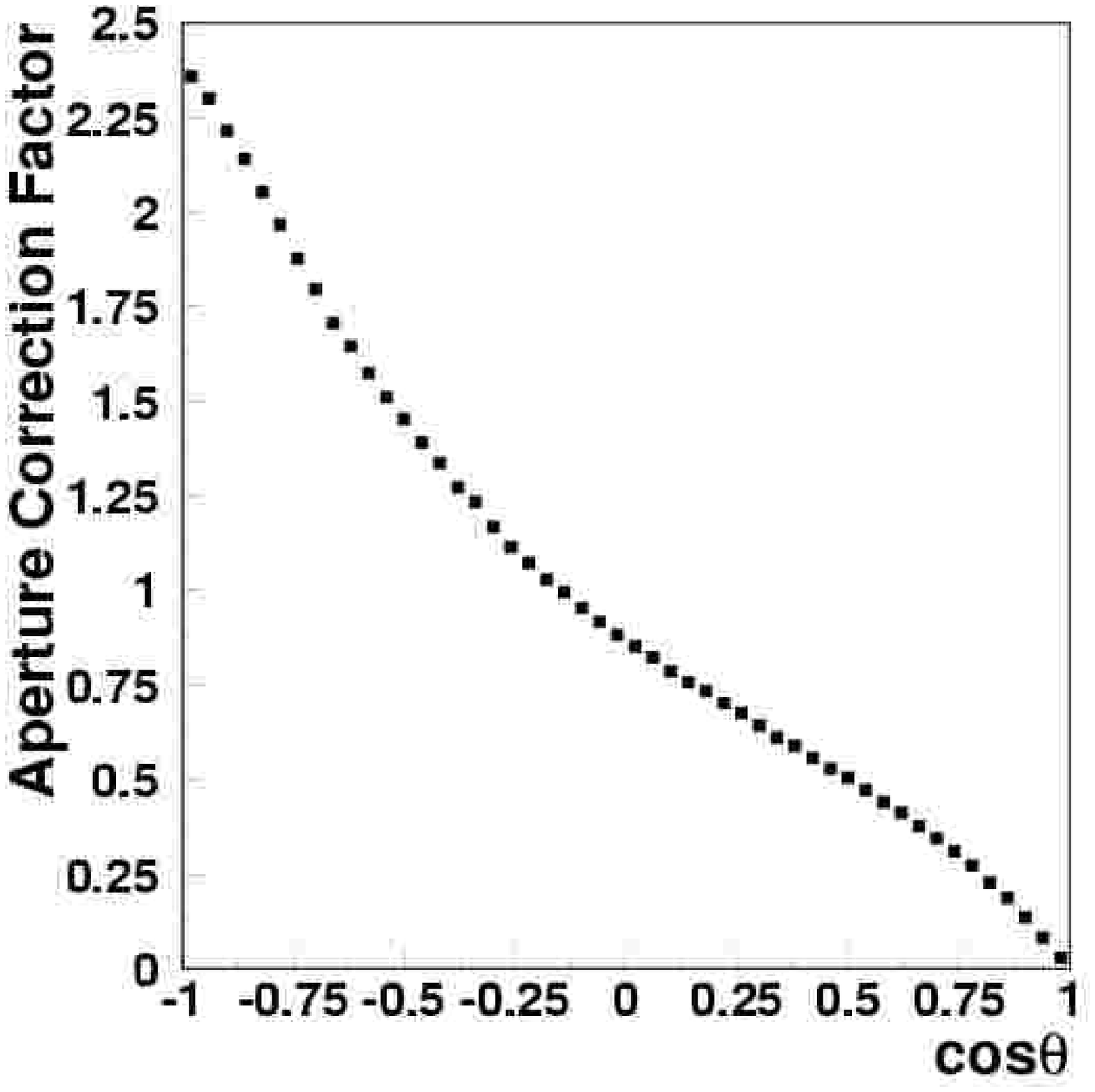}\\
\end{tabular}
\begin{tabular}{c}
(c)\includegraphics[width=6.15cm]{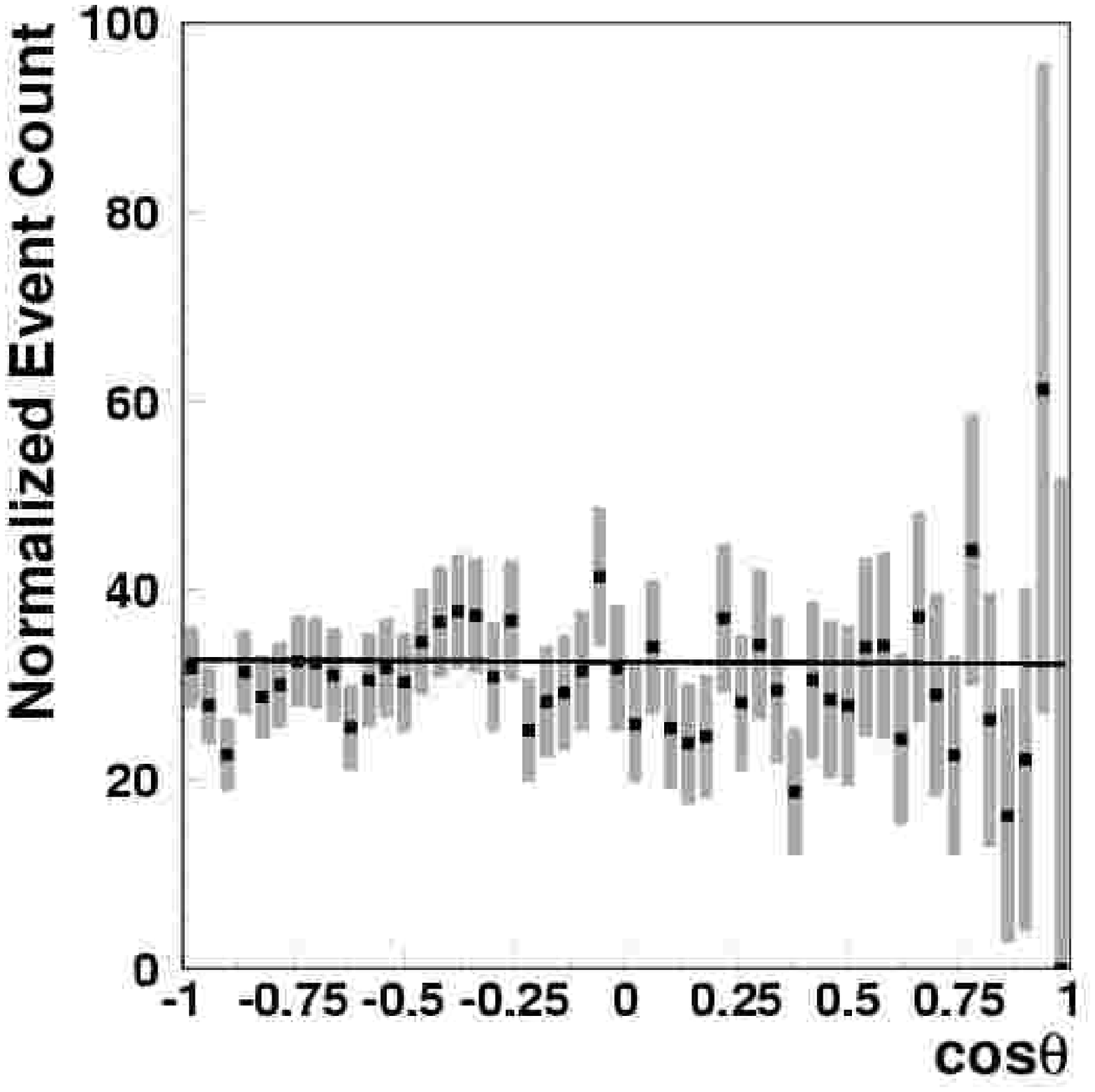}\\
\end{tabular}
\end{center}
\caption{The dipole function for the nominal arrival directions of the 
HiRes-I data set---(a) the number of counts in each $\cos\theta$ bin; 
(b) the aperture/exposure normalization factor for each bin;
(c) the normalized bin count with the $\chi^2$-fit to a line.}
\label{fig:dp_raw}
\end{figure}  

As a first order measurement, we construct the dipole function for a
source model with a maximum value at the Galactic Center.  For now, we only
consider the nominal arrival directions of the events in our data
 sample.  For this demonstration, we set the bin width 
of the dipole function to $\Delta(\cos\theta)=0.04$.  
This provides us with a mean bin count of 30.52.
Figure~\ref{fig:dp_raw}a  shows the resulting dipole function.  
However, in order to estimate the value of $\alpha$, 
we first normalize our
dipole function with respect to aperture and exposure.  This is done by 
considering $10^7$ pairs of simulated events and event times that correspond
to the actual HiRes-I observation periods.  By constructing a dipole function 
for this simulated set, we then estimate the normalization
factor for each $\cos\theta$ bin in the dipole function.  The result is shown
in figure~\ref{fig:dp_raw}b.  The dipole function is then normalized and
a $\chi^2$-fit performed to determine its slope, $m$, and 
$y$-intercept, $b$. The normalized dipole function is
pictured in figure~\ref{fig:dp_raw}c with the best linear fit applied.  
The scaling constant, $\alpha$, is then estimated by the quotient, 
$\frac{m}{b}$.  The result for the 
galactic dipole source model is then: $\alpha=-0.010\pm0.055$.   

The same method
was employed to calculate $\alpha$ in the cases of Centaurus A and M87.  For
Centaurus A, we obtained a result of: $\alpha=-0.035\pm0.060$. For M87, we 
found $\alpha=-0.005\pm0.045$.

\section{Incorporating Angular Resolution into the Dipole Function}

\begin{figure}[t,b,p]
\begin{center}
\begin{tabular}{c}
\includegraphics[width=7.0cm]{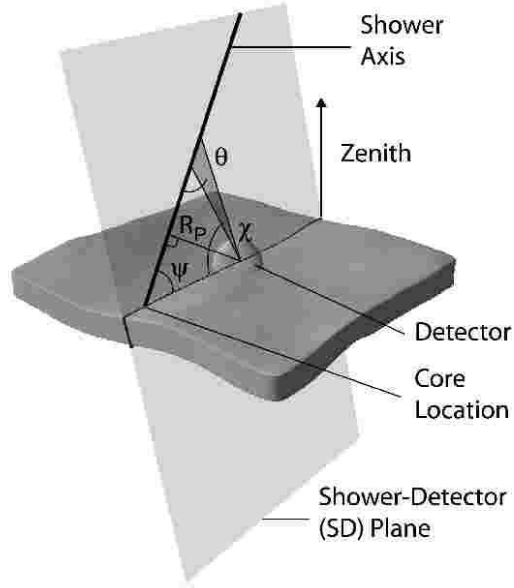}\\
\end{tabular}
\end{center}
\caption{The geometry of reconstruction for a monocular air fluorescence 
detector}
\label{figure:picture}
\end{figure}

The analysis described in the previous section does 
not take into account the experimental angular resolution.  
Accommodating the angular resolution is important to the overall integrity of
this analysis
because the HiRes-I monocular data contains very asymmetric errors 
in arrival direction determination.
For a monocular air fluorescence detector, angular resolution consists
of two components, the error, $\Delta \hat{n}$  
in the estimation of the plane of 
reconstruction and the error, $\Delta\psi$,  
in the determination of the angle, $\psi$, within
the plane of reconstruction.  Figure~\ref{figure:picture} illustrates how
this geometry would appear with a particular plane of reconstruction and a 
particular value for $\psi$.  Intuitively, we can see that the plane of 
reconstruction can be determined quite accurately.  However, the value 
of $\psi$ is more difficult to determine because it is
dependent on the precise results of the profile constraint fit 
\cite{Abu-Zayyad:2002ta,Abu-Zayyad:2002sf}.  In general,
$\Delta \hat{n}$ is dependent upon the observed 
angular track length of the event
in question.  This is because longer track lengths enable a better 
determination of the plane of reconstruction.  Typically, the value of 
$\Delta \hat{n}$ is $\pm0.5^\circ$.  The value of $\Delta\psi$ 
is dependent upon 
the cosmic ray energy.  This is due to the fact
the larger showers provide better defined profiles for the 
reconstruction program.  Typically, the value of $\Delta\psi$ is $\pm10^\circ$.

\begin{figure}[t,b,p]
\begin{center}
\begin{tabular}{c@{\hspace{0.0cm}}c}
(a)\includegraphics[width=6.15cm]{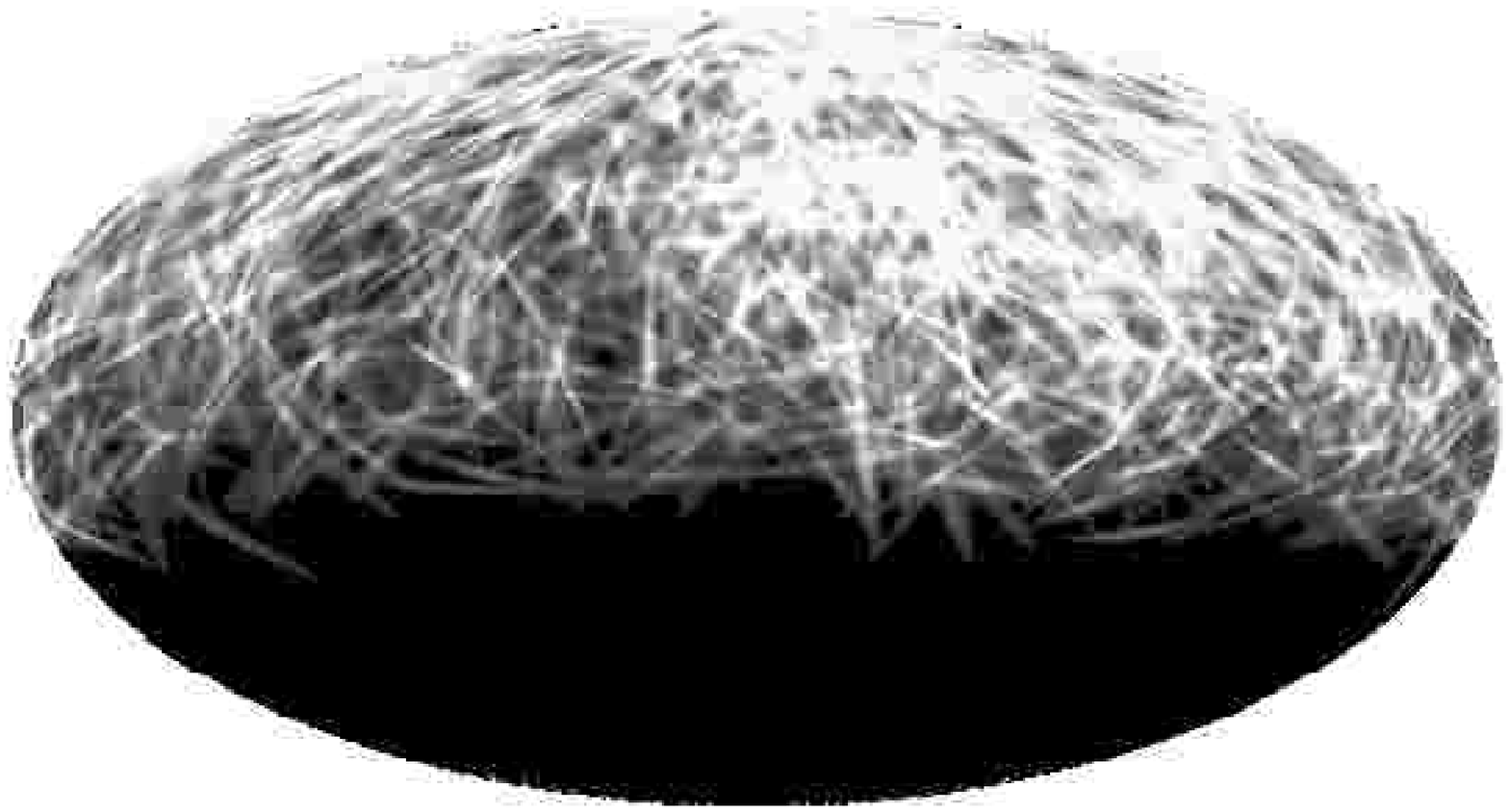}&
(b)\includegraphics[width=6.15cm]{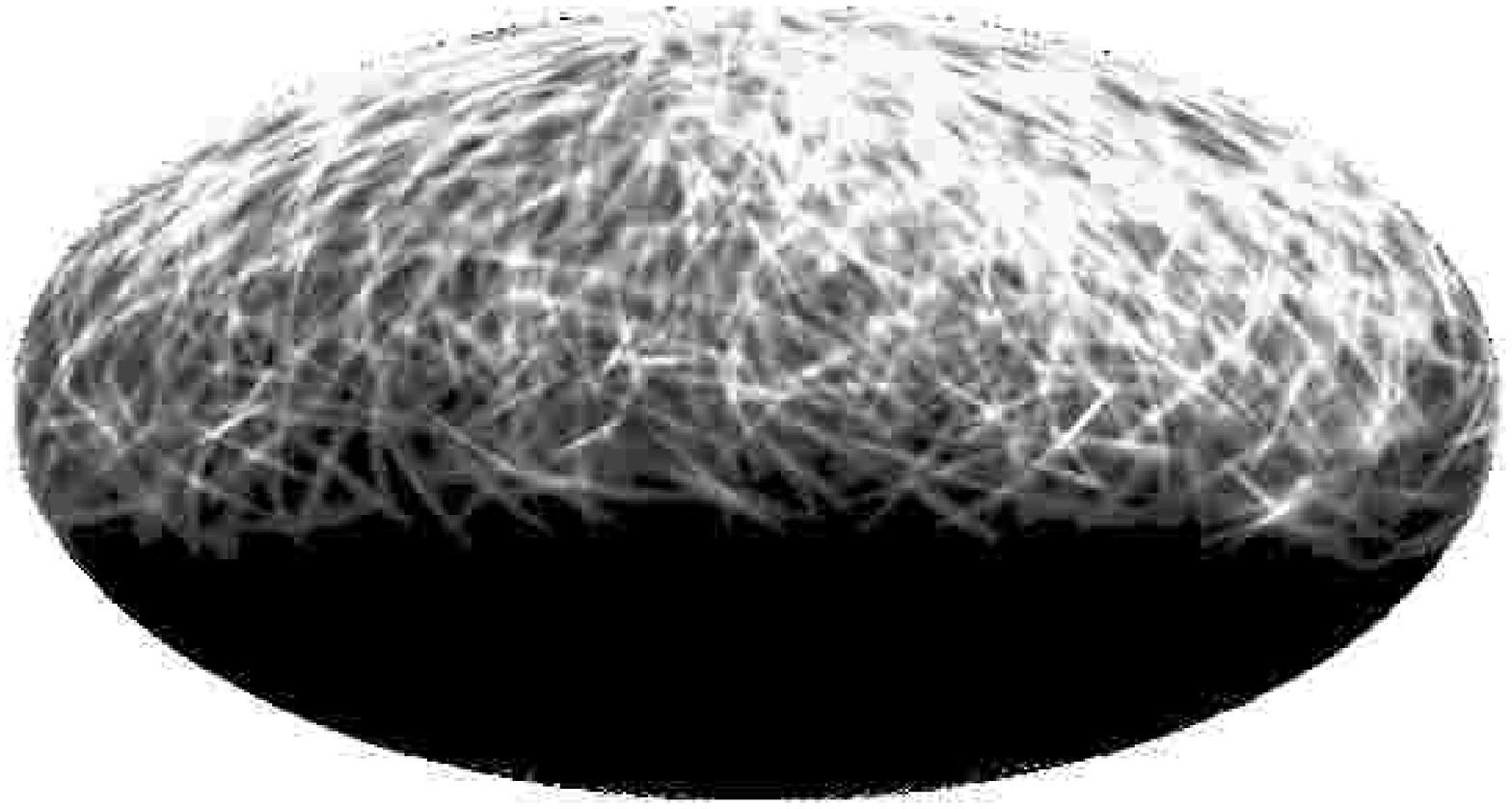}\\
(c)\includegraphics[width=6.15cm]{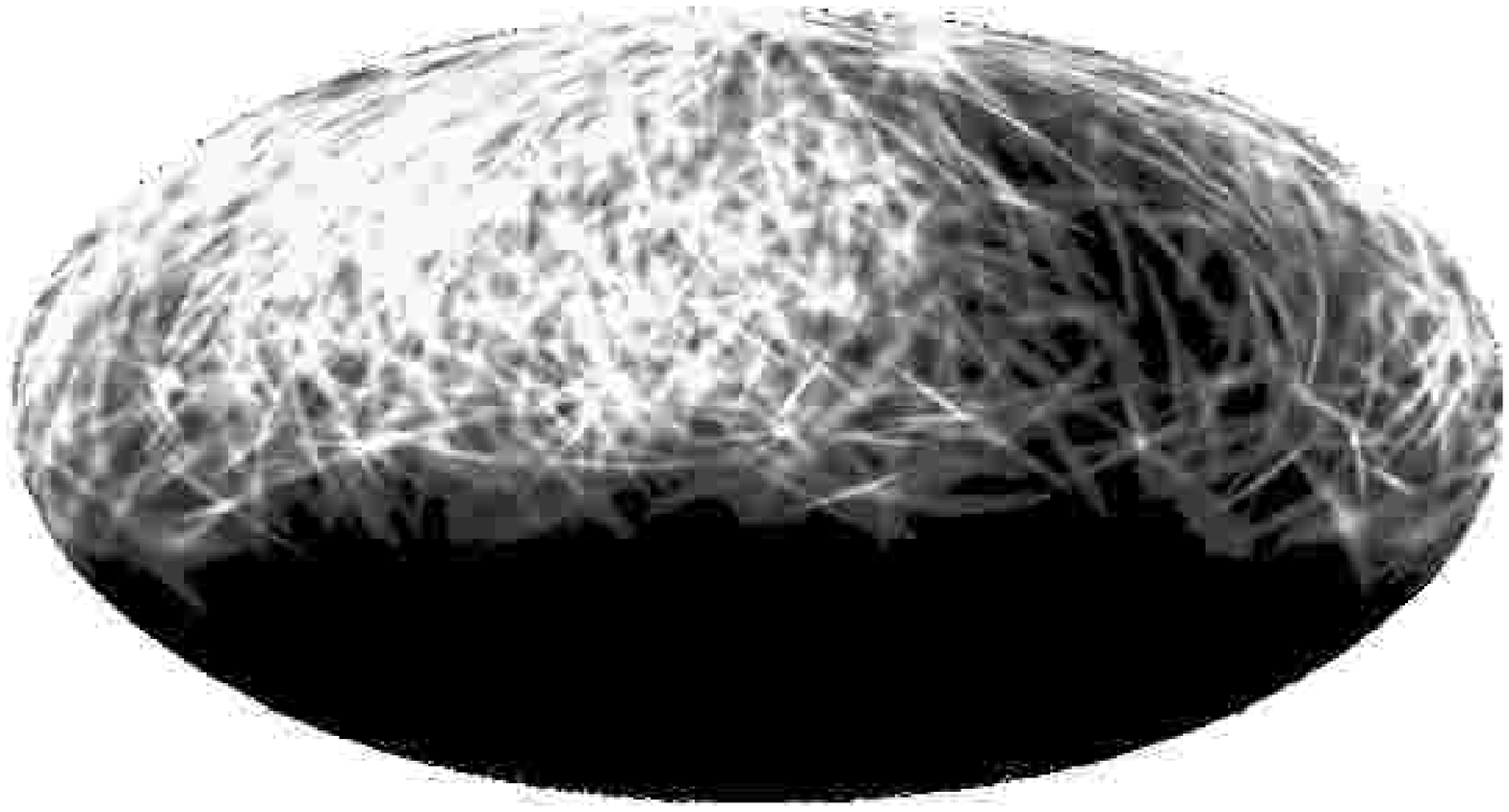}&
(d)\includegraphics[width=6.15cm]{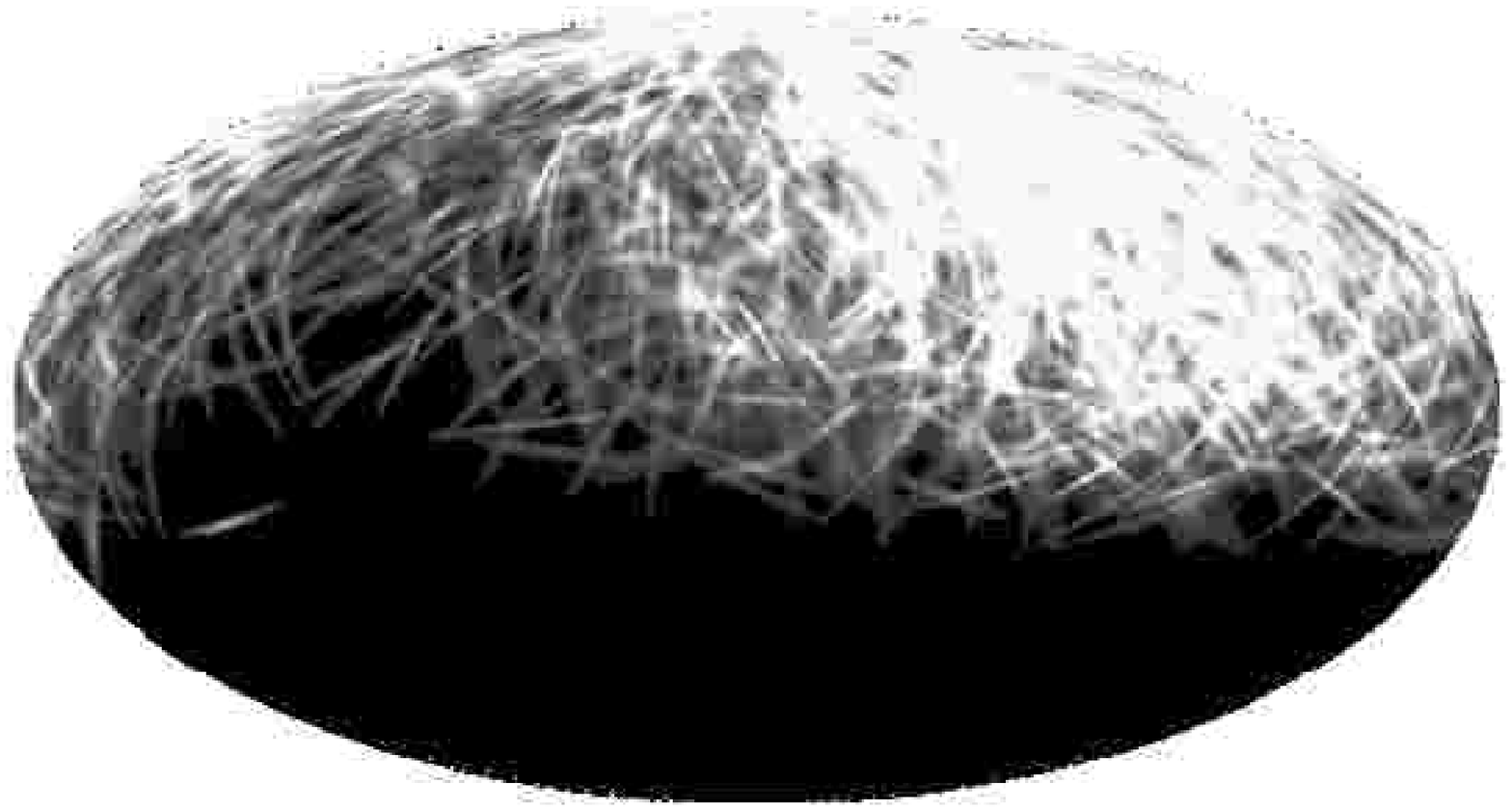}\\
\end{tabular}
\end{center}
\caption{Density plots of event arrival directions  
with the angular resolution parameters of the
Hires-1 monocular data on a Hammer-Aitoff projection with equatorial 
coordinates (right ascension right to left)---(a) 
HiRes-I monocular data set; 
(b) simulated data set with an isotropic source model;
(c) simulated data set with a galactic dipole source model ($\alpha=1$);
(d) simulated data set with a galactic dipole source model ($\alpha=-1$).
In each case, the lighter regions correspond to a higher density of event
arrival directions.}
\label{fig:density}
\end{figure}

In order to accommodate the HiRes-I monocular angular resolution, 
it is necessary to 
revise the method we use to construct the dipole function.  Instead of 
considering each event as a single arrival direction, we will consider
each event to be an elliptical, two-dimensional Gaussian distribution 
of $N$ points 
with the two Gaussian parameters, $\sigma_1$ and $\sigma_2$, being 
defined by the
parameters that describe the angular resolution.  Figure~\ref{fig:density} 
shows how entire sets of events with these error parameters appear when 
projected on a density plot using a Hammer-Aitoff projection and 
equatorial coordinates.

\begin{figure}[t,b,p]
\begin{center}
\begin{tabular}{c@{\hspace{0.0cm}}c}
(a)\includegraphics[width=6.15cm]{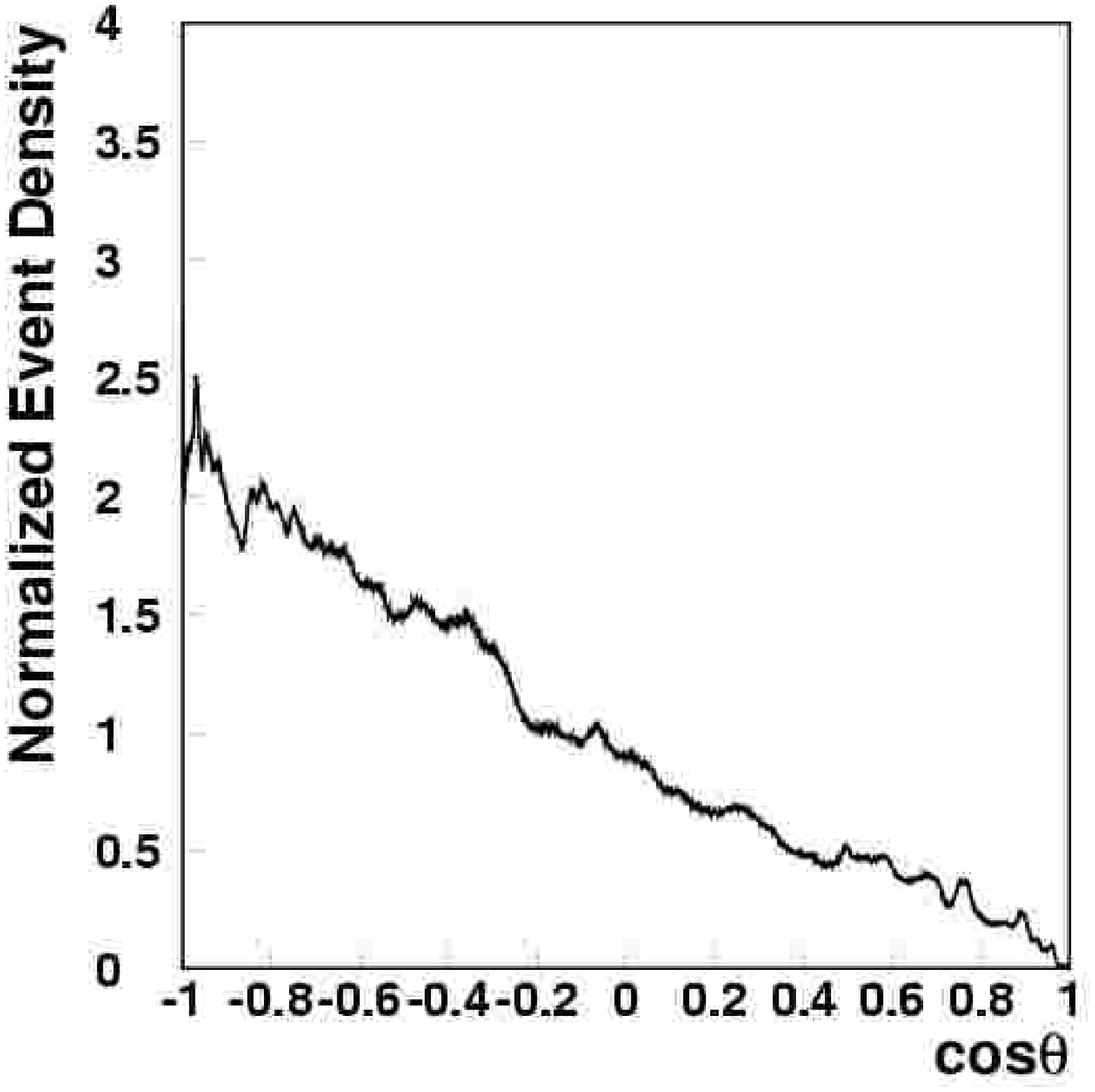}&
(b)\includegraphics[width=6.15cm]{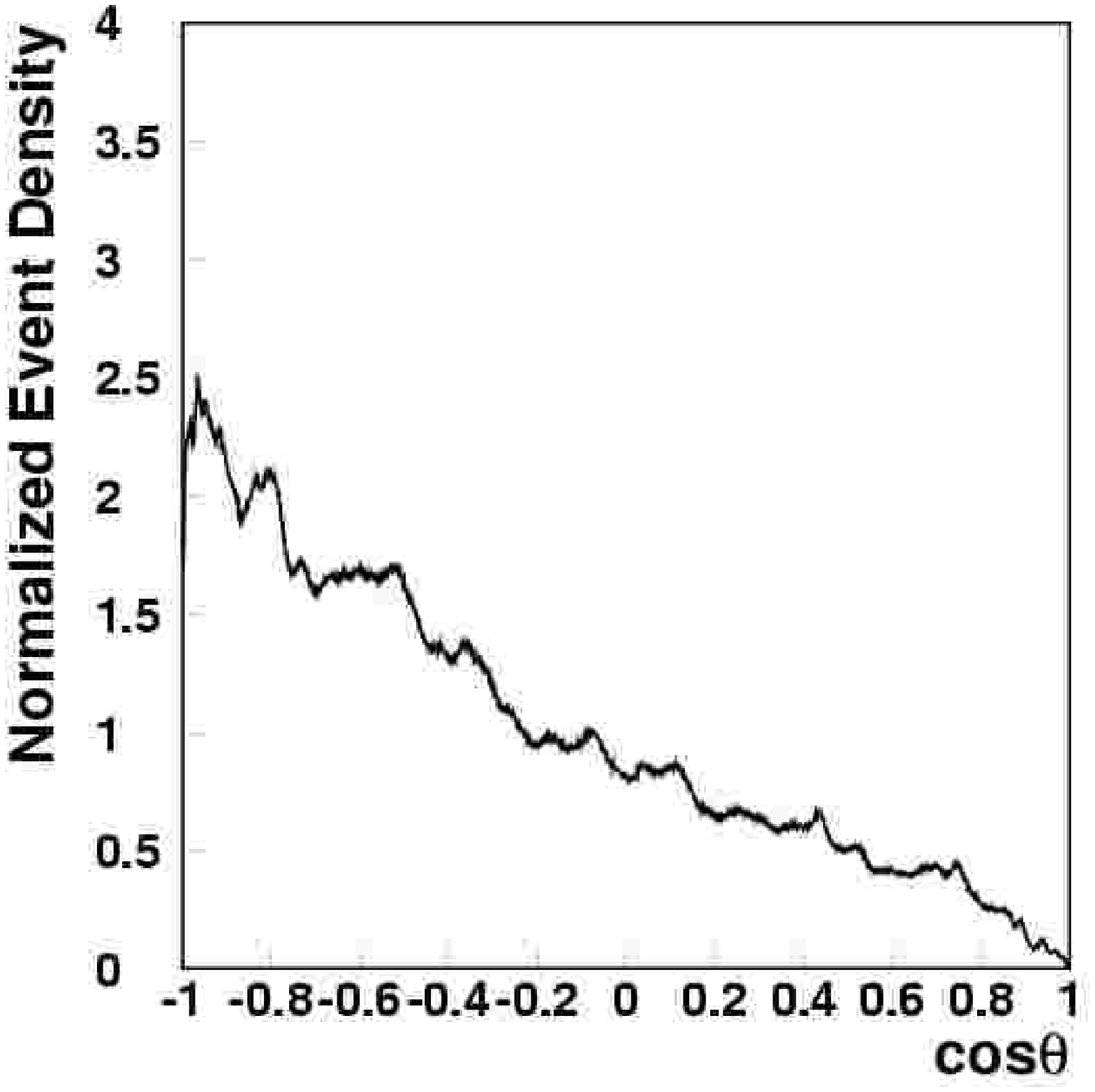}\\
(c)\includegraphics[width=6.15cm]{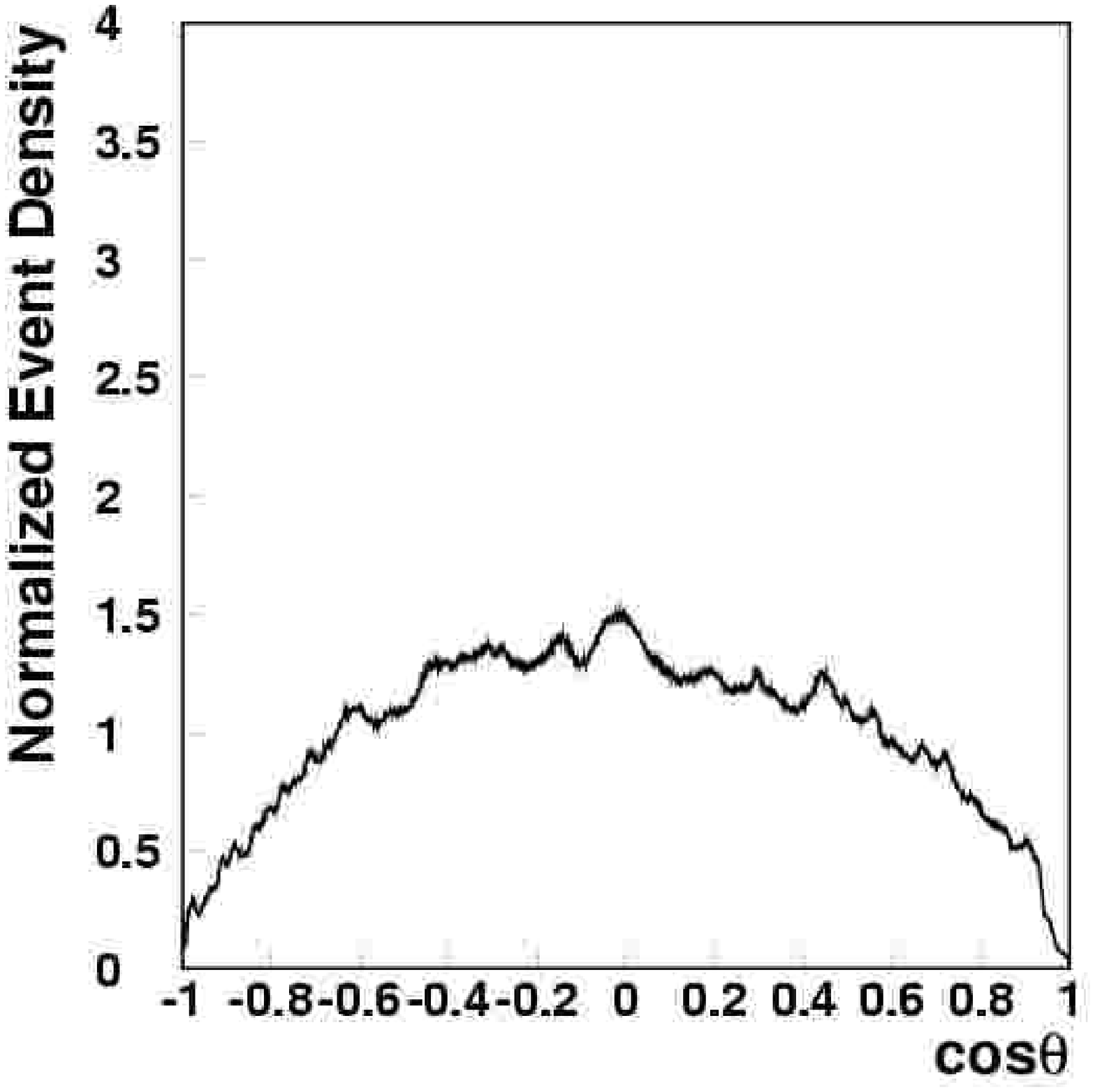}&
(d)\includegraphics[width=6.15cm]{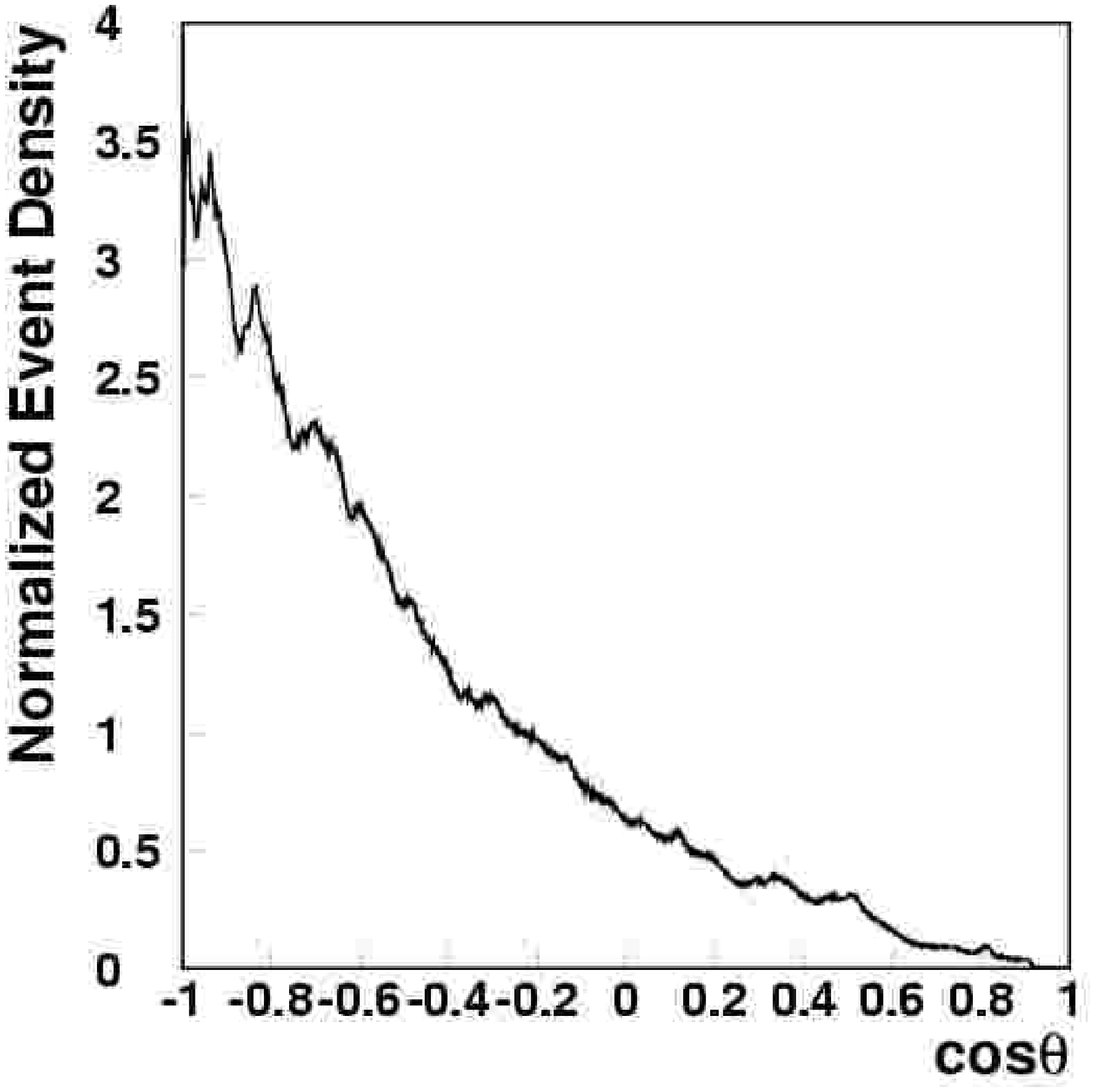}\\
\end{tabular}
\end{center}
\caption{The dipole function, with angular resolution included,
for a galactic dipole model for four different 
event sets without correction for aperture and exposure---(a) 
HiRes-I monocular data set; 
(b) simulated data set with an isotropic source model;
(c) simulated data set with a galactic dipole source model ($\alpha=1$);
(d) simulated data set with a galactic dipole source model ($\alpha=-1$).}
\label{fig:dpf}
\end{figure}

In order to account for angular resolution in the construction of the dipole 
function, we add an additional step.  Instead of simply calculating
the opening angle between the arrival direction of the event and the center
of the dipole for the preferred arrival direction, we do so 
separately for each of the $N$ points 
in the Gaussian distribution that describes each event's arrival direction.  
By choosing a sufficiently large value for $N$ and a sufficiently small bin 
width, $\Delta(\cos\theta)$, 
 we can then construct the dipole function as a smooth curve.
Examples of the dipole function 
are shown in figure~\ref{fig:dpf} for each of the four event sets in
figure~\ref{fig:density}.

The next logical step would be to attempt to normalize the dipole function
of the real data with respect to aperture and exposure and then to calculate 
the slope, $m$, and the $y$-intercept, $b$.  
However, this program would run into a major complication.  Because
the Gaussian distributions that are used to approximate the individual event
arrival directions can overlap into a large number of bins, 
the individual data points in 
the dipole function are highly correlated.  This makes it impossible to apply
either the $\chi^2$-fit or a bootstrap method to estimate 
the error in the values of $m$ and $b$---and thus the error in 
$\alpha$---for the normalized dipole function. 
Another approach needs to be developed.

The method that we propose is to compare the value of $<\!\!\cos\theta\!\!>$ for the 
dipole function of the real data sample with that of a large number of
similar-sized simulated data samples  
with a discrete spectrum of $\alpha$-values.  We can then 
show how $<\!\!\cos\theta\!\!>$ varies with 
respect to $\alpha$ for different dipole source models.   

\section{Simulating the HiRes Aperture and Exposure}

In creating simulated data sets, we employed a library  
of simulated events generated by our Monte Carlo shower simulation
program and then  
reconstructed using the profile-constraint reconstruction program.  This 
library of events possesses the spectrum and composition reported
by the stereo Fly's Eye experiment\cite{Bird:wp,Bird:comp}.  
A total of $\sim1.3\times10^{5}$ 
simulated events were reconstructed with energies greater than 
$10^{18.5}$~eV.  

Once a library of simulated events was created, we then turned to the task 
of creating simulated data sets that accurately reflected the exposure of the
HiRes-I monocular data set.  
In general, the apertures of air-fluorescence detectors 
are complicated; we need to assign times to individual Monte Carlo events
that accurately reflect the distribution of  
times seen in the actual data. 

By parsing through the raw HiRes-I data, we assemble a database of detector
run-periods.  
We then randomly assign a time from these periods to each simulated 
event for a simulated event set.  We also apply a further correction to 
account for the effect of non-functioning detector units (mirrors).  This is
achieved by excluding mirror events corresponding to periods in which a
particular mirror was out of commission.

\begin{figure}[t,b,p]
\begin{center}
\begin{tabular}{c}
\includegraphics[width=8.0cm]{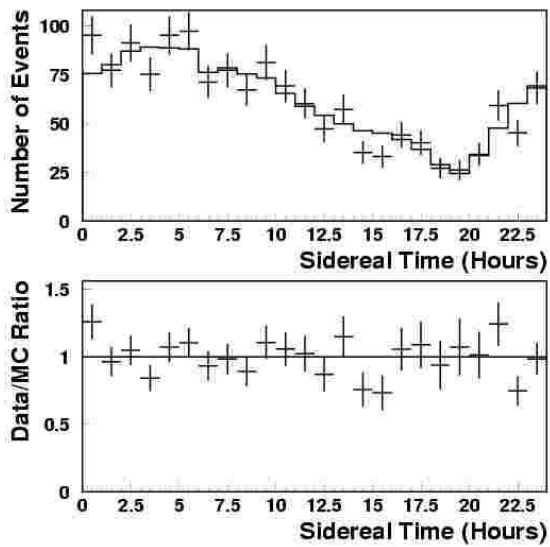}\\
\end{tabular}
\end{center}
\caption{Sidereal time distribution comparison between the real data and 
a simulated data after the mirror-by-mirror correction ($\chi^2/d.f.=1.18$).  
The solid line histogram corresponds to the 
sidereal time distribution of the simulated data.  The crosses 
correspond to the sidereal time distribution of the real data with Gaussian
uncertainties assumed for each bin.}
\label{figure:sidereal}
\end{figure}

In figure~\ref{figure:sidereal} we can see the results of this 
mirror-by-mirror correction by comparing the sidereal time distributions of 
real and simulated data sets after the correction is applied.  We see 
excellent agreement in this plot ($\chi^2/d.f.=1.18$).

\begin{figure}[t,b,p]
\begin{center}
\begin{tabular}{c}
\includegraphics[width=8.0cm]{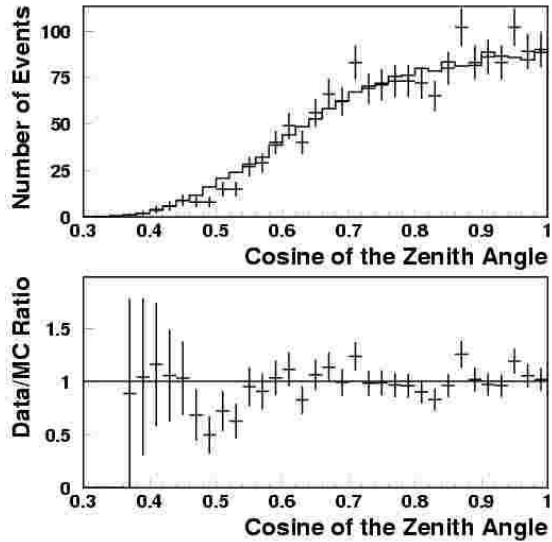}\\
\end{tabular}
\end{center}
\caption{Zenith angle distribution comparison between the real data and 
simulated data ($\chi^2/d.f.=0.93$).  
The solid line histogram corresponds to the 
distribution of cosine of the zenith angles for the simulated data.  
The crosses correspond to the distribution of cosine of the zenith angles for
 the real data with Gaussian uncertainties assumed for each bin.}
\label{figure:zen}
\end{figure}

\begin{figure}[t,b,p]
\begin{center}
\begin{tabular}{c}
\includegraphics[width=8.0cm]{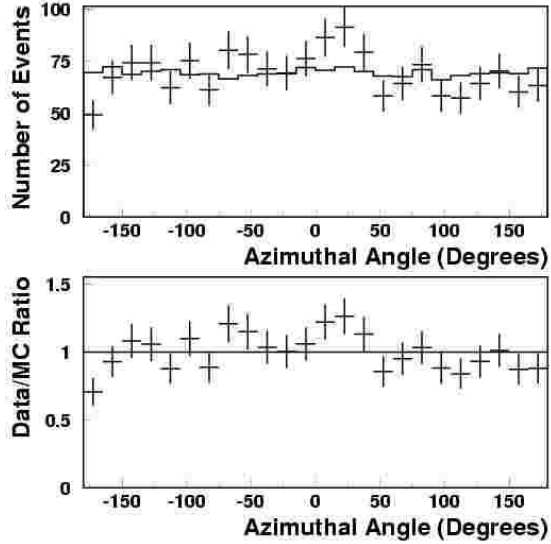}\\
\end{tabular}
\end{center}
\caption{Azimuth angle distribution comparison between the real data and 
simulated data ($\chi^2/d.f.=1.31$).  The solid line histogram 
corresponds to the distribution azimuth angles for the simulated data. 
The crosses correspond to the distribution of azimuth angles for the real data
 with Gaussian uncertainties assumed for each bin.}
\label{figure:azi}
\end{figure}

We also checked to see if the Monte Carlo shower simulation
routine was accurately modeling
the efficiency of the HiRes-I detector with respect to zenith and azimuth
angles.  In figures~\ref{figure:zen}~and~\ref{figure:azi}, 
we compare the distributions of zenith 
and azimuth angles for the real data and the simulated data set
that has been assigned random times and filtered through our 
mirror-by-mirror correction.  There is again very good agreement between the 
simulation and the data.

\section{Results}

\begin{figure}[t,b,p]
\begin{center}
\begin{tabular}{c@{\hspace{0.0cm}}c}
(a)\includegraphics[width=6.15cm]{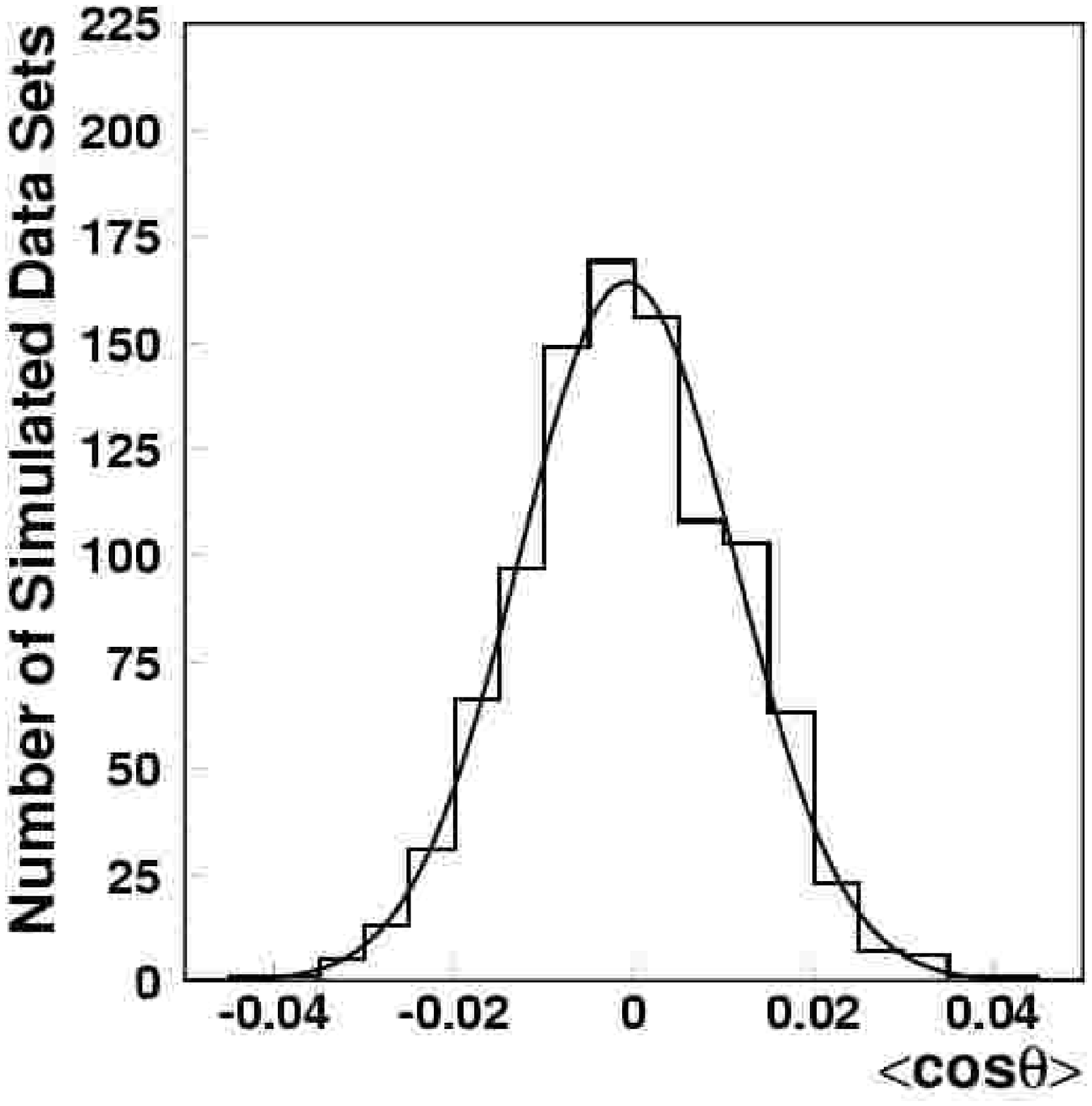}&
(b)\includegraphics[width=6.15cm]{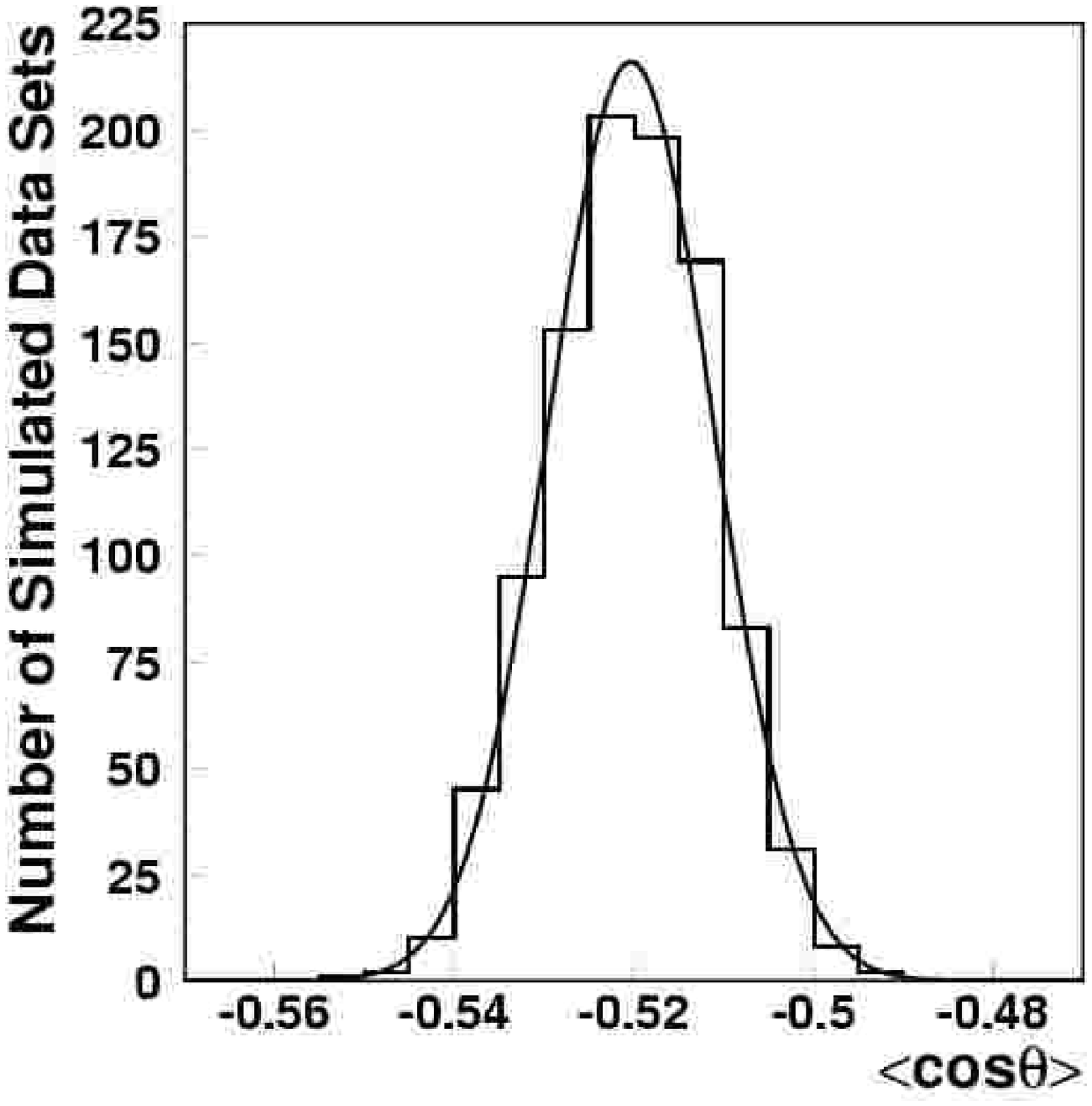}\\
\end{tabular}
\end{center}
\caption{The distribution of $<\!\!\cos\theta\!\!>$ values for the dipole functions
of simulated data sets with a single $\alpha$-value---(a) 
the galactic dipole source model with $\alpha=1.0$; 
(b) the galactic dipole source model with $\alpha=-1.0$.}
\label{fig:dpmd}
\end{figure}

For each of the three dipole source models considered we used the following
procedure to measure the $\alpha$ parameter:
\begin{enumerate}
\item We calculated the value of $<\!\!\cos\theta\!\!>$ for the dipole function of the real data sample.  
\item We created a total of 20,000 simulated data samples, 1000 each for 
0.1 increments of $\alpha$ from -1.0 to 1.0, each with the same number of 
events as the actual data.  In figure~\ref{fig:dpmd} we can 
see that the distribution of $<~\!\!\!\!\cos\theta\!\!>$ 
values for each $\alpha$-value
generated conforms well to a Gaussian distribution.  
\item We constructed curves corresponding to the mean and standard
deviation of $<\!\!\cos\theta\!\!>$ of the dipole function for each value of $\alpha$. 
\item We determined the preferred value of $\alpha$ and the 90\% 
confidence interval of $\alpha$ for each dipole source model by referring to 
the intersections of the 90\% confidence interval curves with the actual value
of $<\!\!\cos\theta\!\!>$ for the dipole function of the real data.
\end{enumerate}

\begin{figure}[t,b,p]
\begin{center}
\begin{tabular}{c@{\hspace{0.0cm}}c}
(a)\includegraphics[width=6.15cm,height=5.70cm]{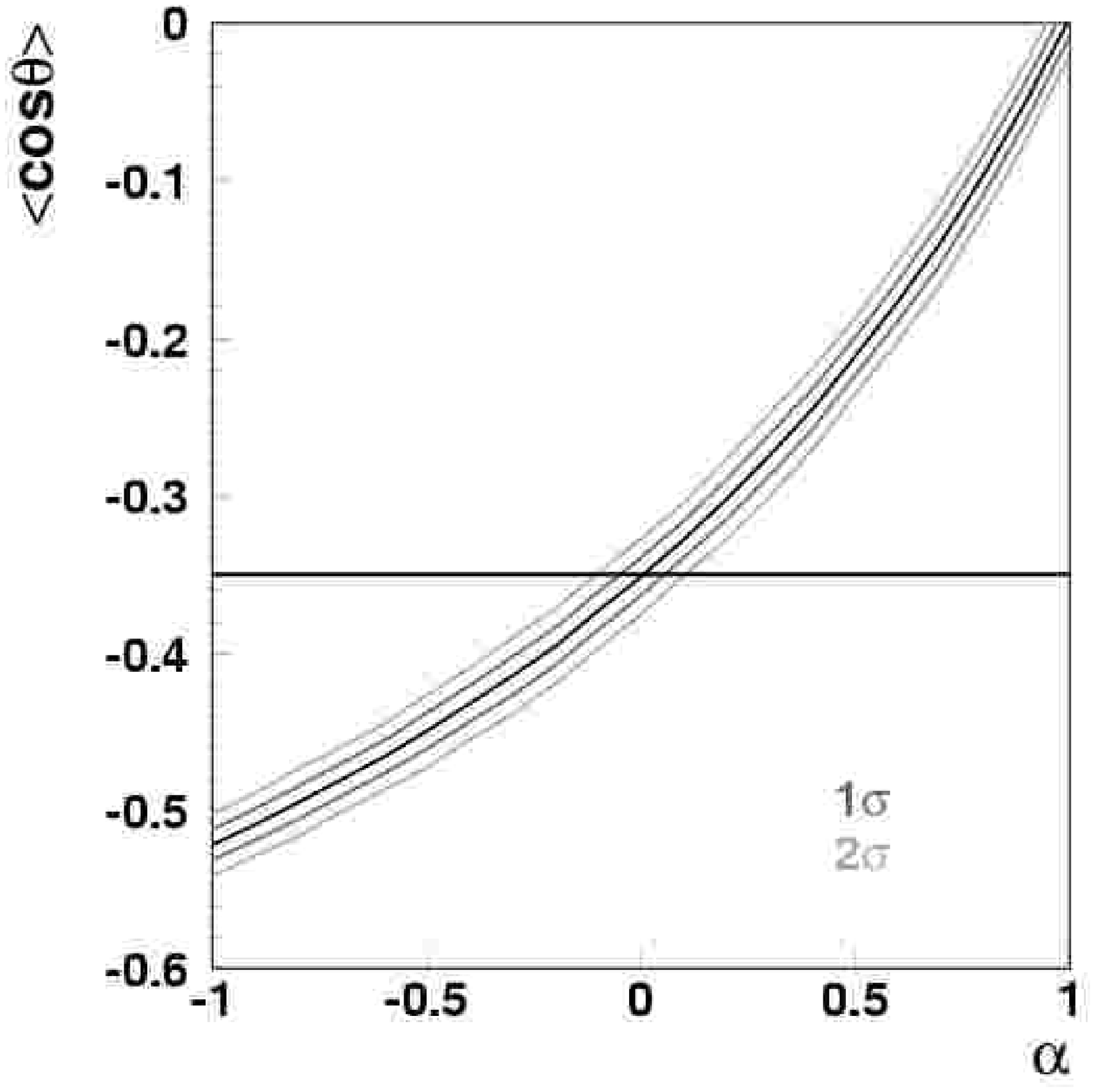}&
(b)\includegraphics[width=6.15cm,height=5.70cm]{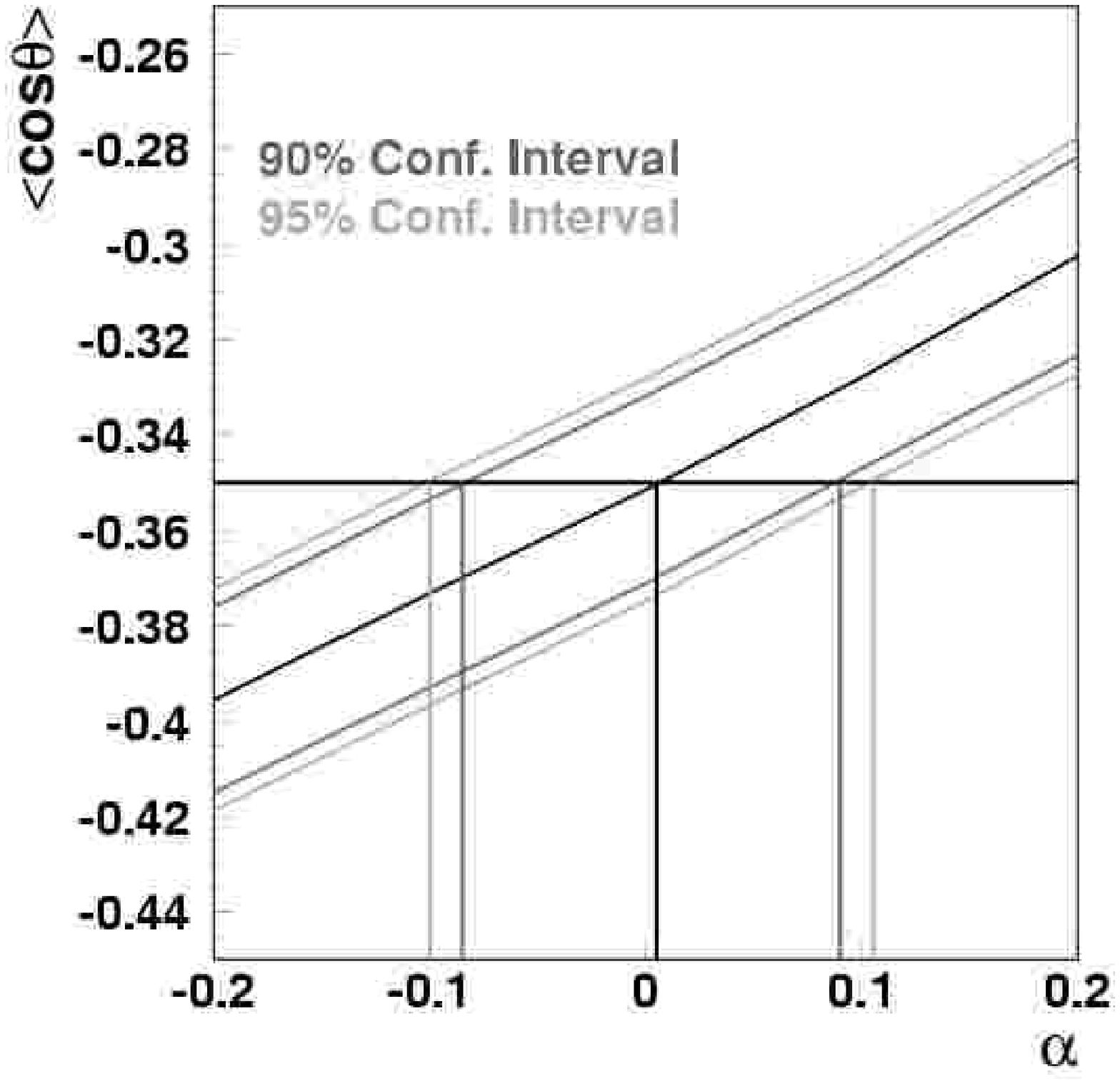}\\
(c)\includegraphics[width=6.15cm,height=5.70cm]{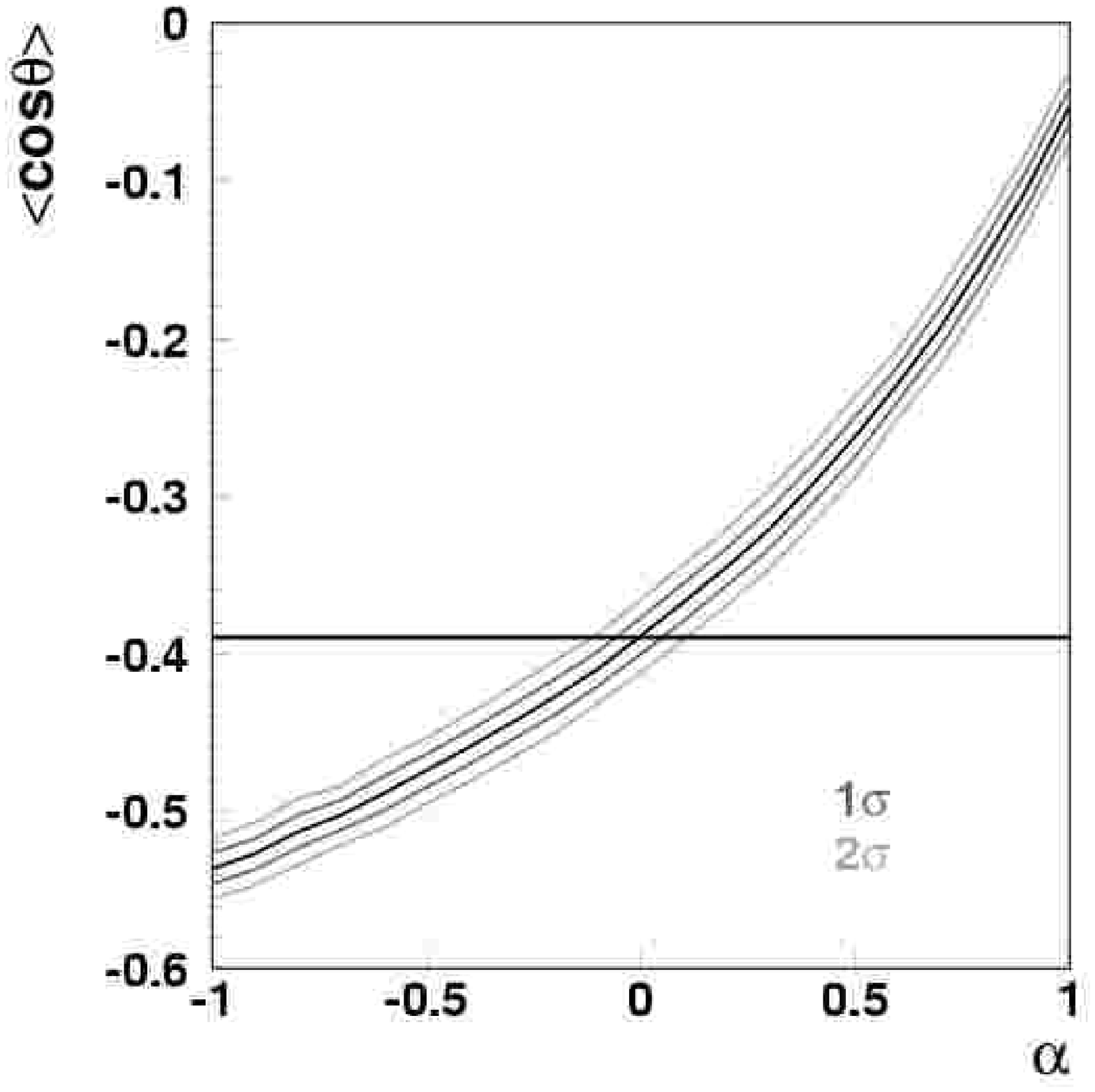}&
(d)\includegraphics[width=6.15cm,height=5.70cm]{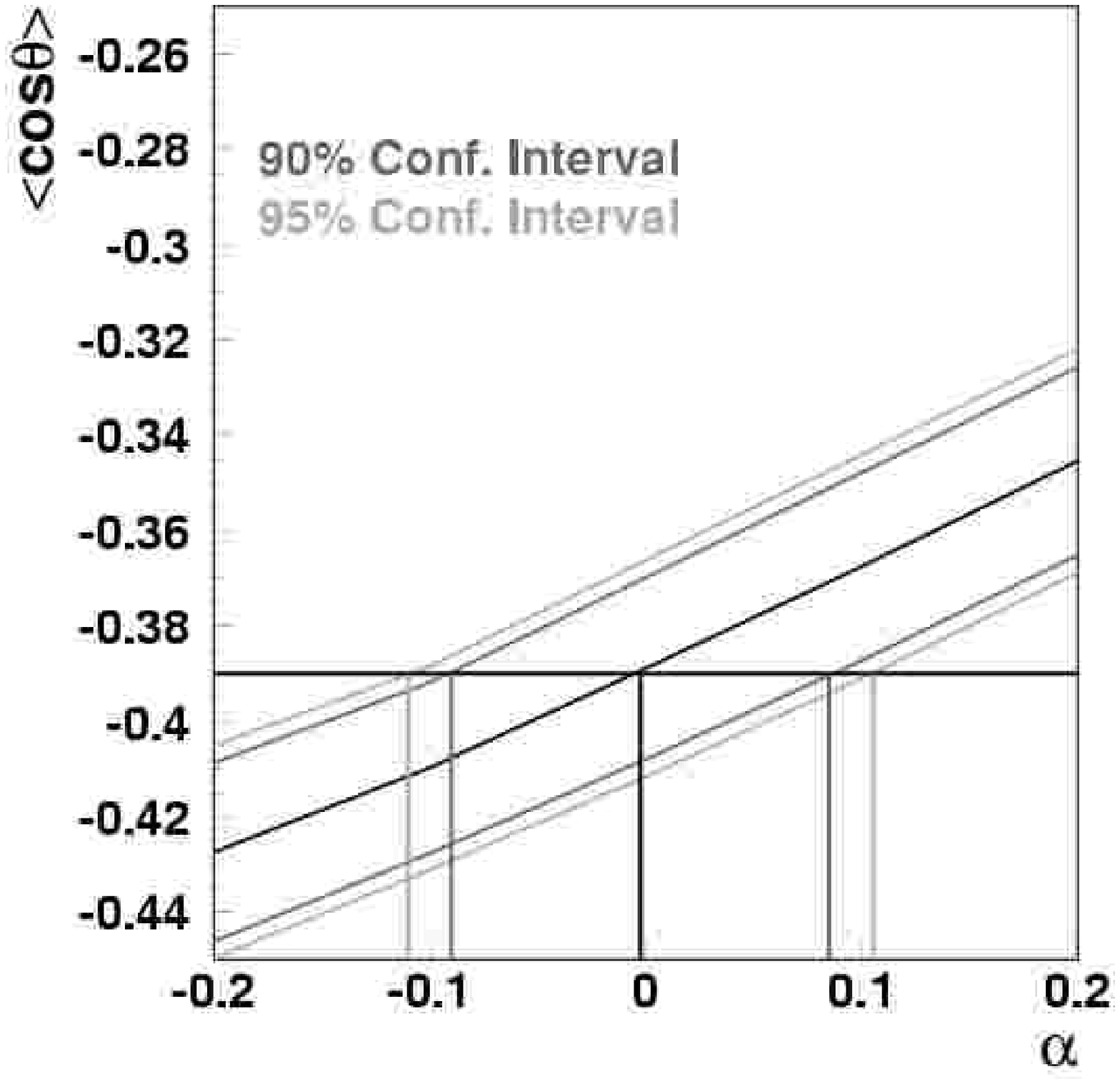}\\
(e)\includegraphics[width=6.15cm,height=5.70cm]{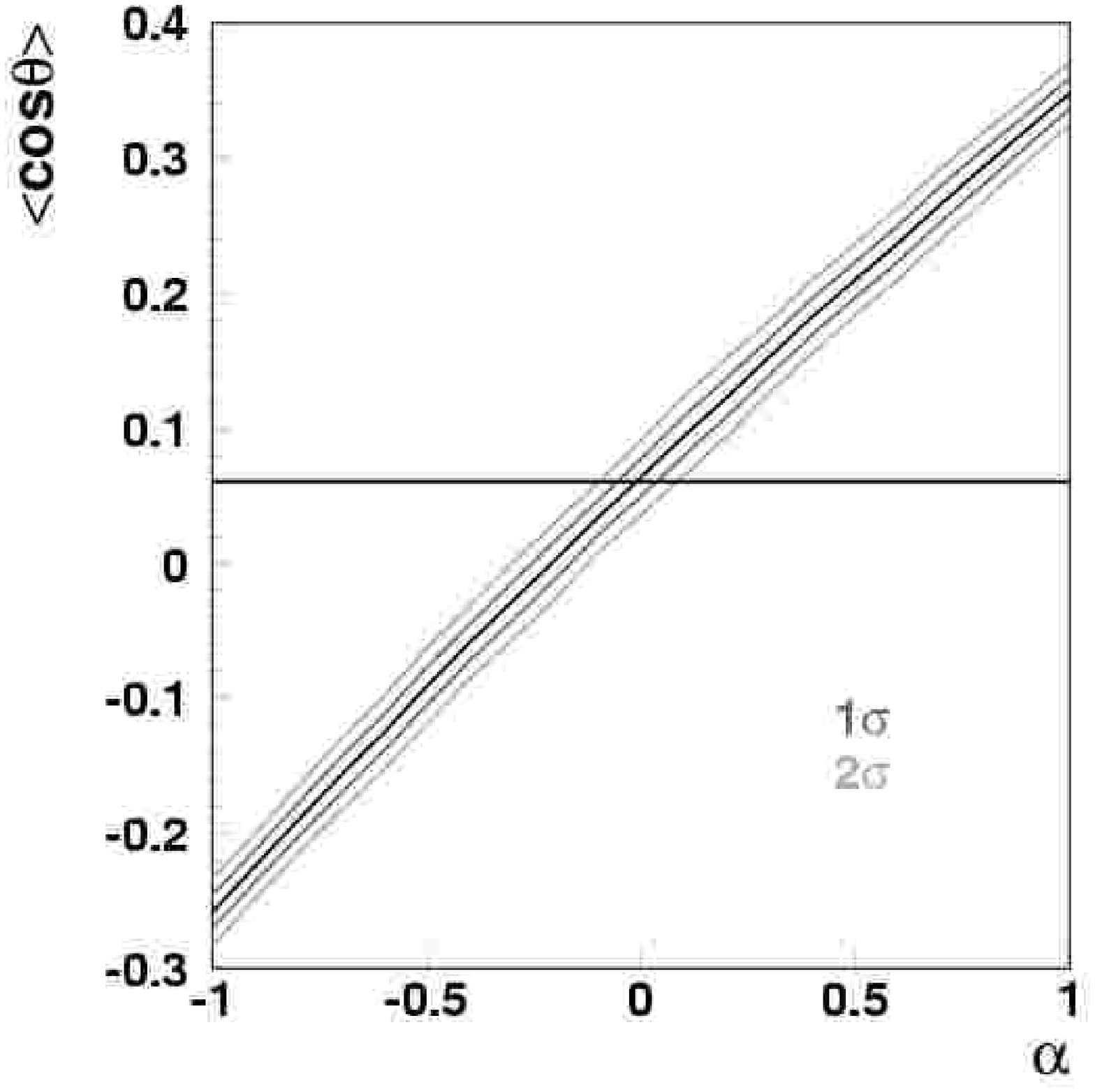}&
(f)\includegraphics[width=6.15cm,height=5.70cm]{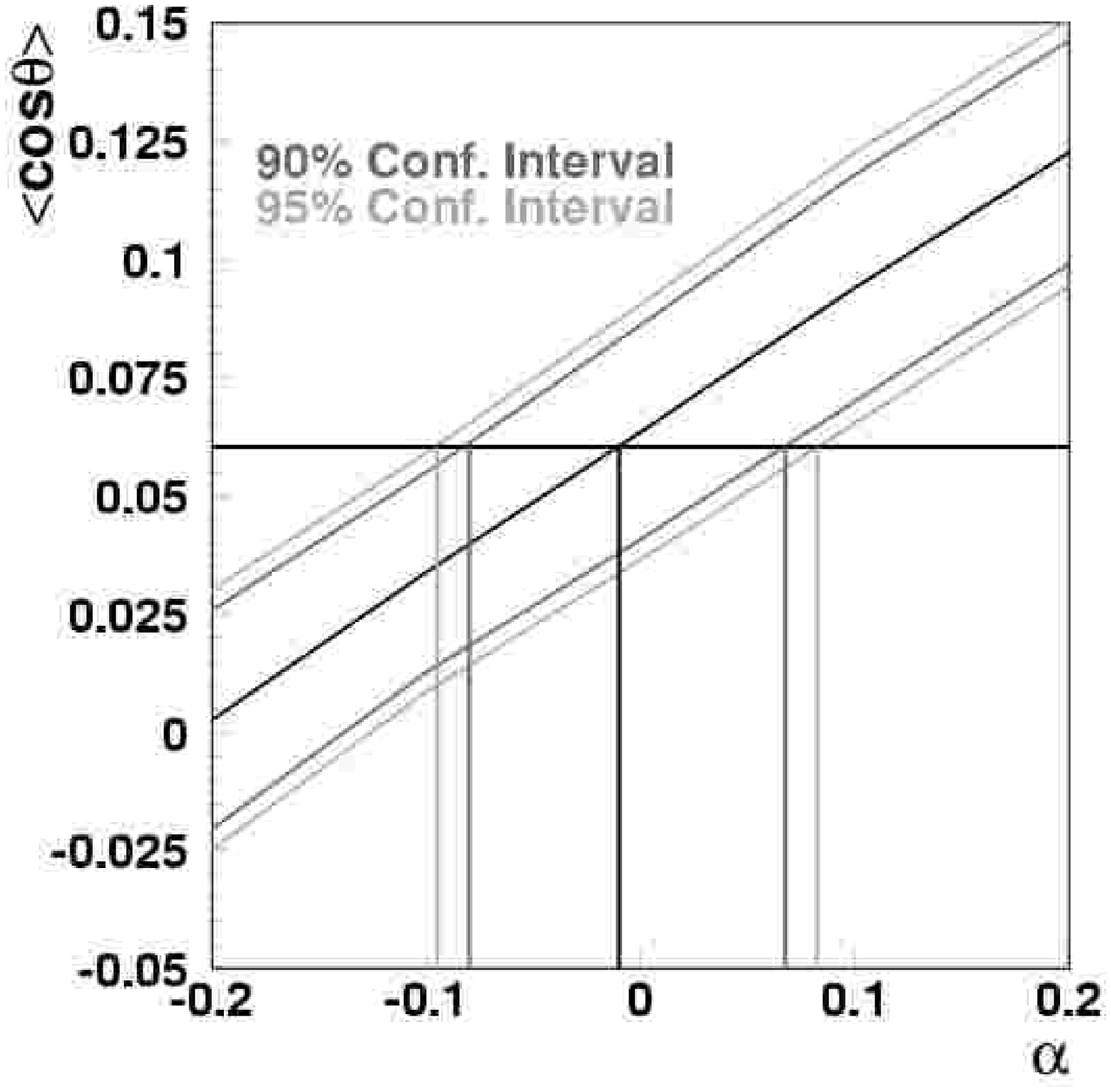}\\
\end{tabular}
\end{center}
\caption{Estimations of the value of $\alpha$ for three different dipole
source models.  The curves demonstrate the dependence of $<\!\!\cos\theta\!\!>$ of 
the dipole functions upon 
$\alpha$.  The horizontal lines represent the value of $<\!\!\cos\theta\!\!>$ of
the real data for the dipole functions of each dipole source model---(a) 
the galactic dipole source model for $\alpha=[-1.0,1.0]$;
(b) the critical region for the galactic dipole model: 
$\alpha=0.005\pm0.055$ with a 90\% confidence interval of: $[-0.085,0.090]$;
(c) the Centaurus A dipole source model for $\alpha=[-1.0,1.0]$;
(d) the critical region for the Centaurus A dipole model: 
$\alpha=-0.005\pm0.065$ with a 90\% confidence interval of: $[-0.090,0.085]$; 
(e) the M87 dipole source model for $\alpha=[-1.0,1.0]$;
(f) the critical region for the M87 dipole model: $\alpha=-0.010\pm0.045$ with
a 90\% confidence interval of: $[-0.080,0.070]$. 
}
\label{fig:dpm}
\end{figure}

The results for all three dipole source models are shown in 
figure~\ref{fig:dpm}. In each case, the nominal values of $\alpha$ and the
90\% confidence levels only deviated marginally from the values obtained 
without considering angular resolution.  The results are given in column~2 of
table~\ref{table:comp}.
 
\section{Potential Sources of Systematic Error in the Estimation of $\alpha$}

There are two principal potential sources of systematic error in the 
determination of $\alpha$ with HiRes-I monocular data.  The first lies in the 
estimation of the angular resolution.  If 
the error in arrival direction estimation was being underestimated or
overestimated, it could lead to an improper evaluation 
of  the confidence intervals
for $\alpha$.  In order to study the effect of angular resolution on our
determination of $\alpha$, we repeated our analysis of the
galactic dipole model twice.  In the first case, we increased the estimated
angular resolution parameters for both the real and simulated data sets by 
33\%.  In the second  case, we decreased the angular resolution parameters for
both types of data sets by 25\%.  In both cases, the width of the 90\% 
confidence interval for $\alpha$ changed by less than 0.010 and the nominal 
value of $\alpha$ remained
unchanged. The results suggest that the determination of $\alpha$
is largely independent of the angular resolution---at least for the plausible
range of values that one could adopt for the angular resolution parameters.

The second issue of concern is the uncertainty in the 
determination of atmospheric clarity.  Because hourly atmospheric observations
are not available for the entire HiRes-I monocular data set, we
have relied upon the use of an average atmospheric profile for the 
reconstruction of our data \cite{Wiencke:atmos}.  
Different atmospheric conditions can influence 
how the profile constraint reconstruction routine interprets an observed 
shower profile and thus can lead to slightly divergent determinations of an
event's arrival direction.  Unfortunately, we do not have large
libraries of simulated data with differing atmospheric parameters used in 
the generation and reconstruction of events.  
However, we do have the real data 
reconstructed with a full range of atmospheric parameters.  By considering 
the value of $<\!\!\cos\theta\!\!>$ over 
the $1\sigma$ error space of atmospheric 
parameters, we can establish the degree of systematic uncertainty that is
contributed to the determination of $\alpha$ by atmospheric variability.
We saw that in the most extreme case, the nominal
value of $\alpha$ shifted by less than $.01$.  
There was no broadening in the 90\% confidence interval.   

\section{Using the Information Dimension, $D_{\rm I}$, as an Independent Check}

The information dimension, $D_{\rm I}$ \cite{balatoni,nayfeh}, 
is a measure of the overall heterogeneity of a data sample.  The smaller the
value of $D_{\rm I}$, the more heterogeneous the sample is.  A basic formula
for calculating $D_{\rm I}$ is:
\begin{equation}
D_{\rm I}=\biggr<\!\! -\frac{1}{\log N_{\rm \delta}}\sum_{i=1}^{N} 
P_{\rm i}(N_{\rm \delta})\log P_{\rm i}(N_{\rm \delta})\!\biggr>,\: 
N_{\rm\delta}=[354,360], 
\end{equation}
where $N_{\rm\delta}$ is the total number of declinational bins (with a range
of values between 354 to 360) and:
\begin{equation}
P_{\rm i}(N_{\rm\delta})=\frac{n_{\rm i}}{<\!\! n_{\rm i}\!\!>}\frac{\pi^{3}}
{4(N_{\rm\delta})^{4}\Delta\Omega_{\rm\delta}},
\end{equation}
with $n_{\rm i}$ being the number of counts in a particular latitudinal bin, 
$<\!\!n_{\rm i}\!\!>$ being the average bin count over the entire sample and 
$\Delta\Omega_{\rm\delta}$ being the area of that particular latitudinal bin.
A detailed description of this method can be found in reference \cite{stokes}.

\begin{figure}[t,b,p]
\begin{center}
\begin{tabular}{c@{\hspace{0.0cm}}c}
(a)\includegraphics[width=6.15cm]{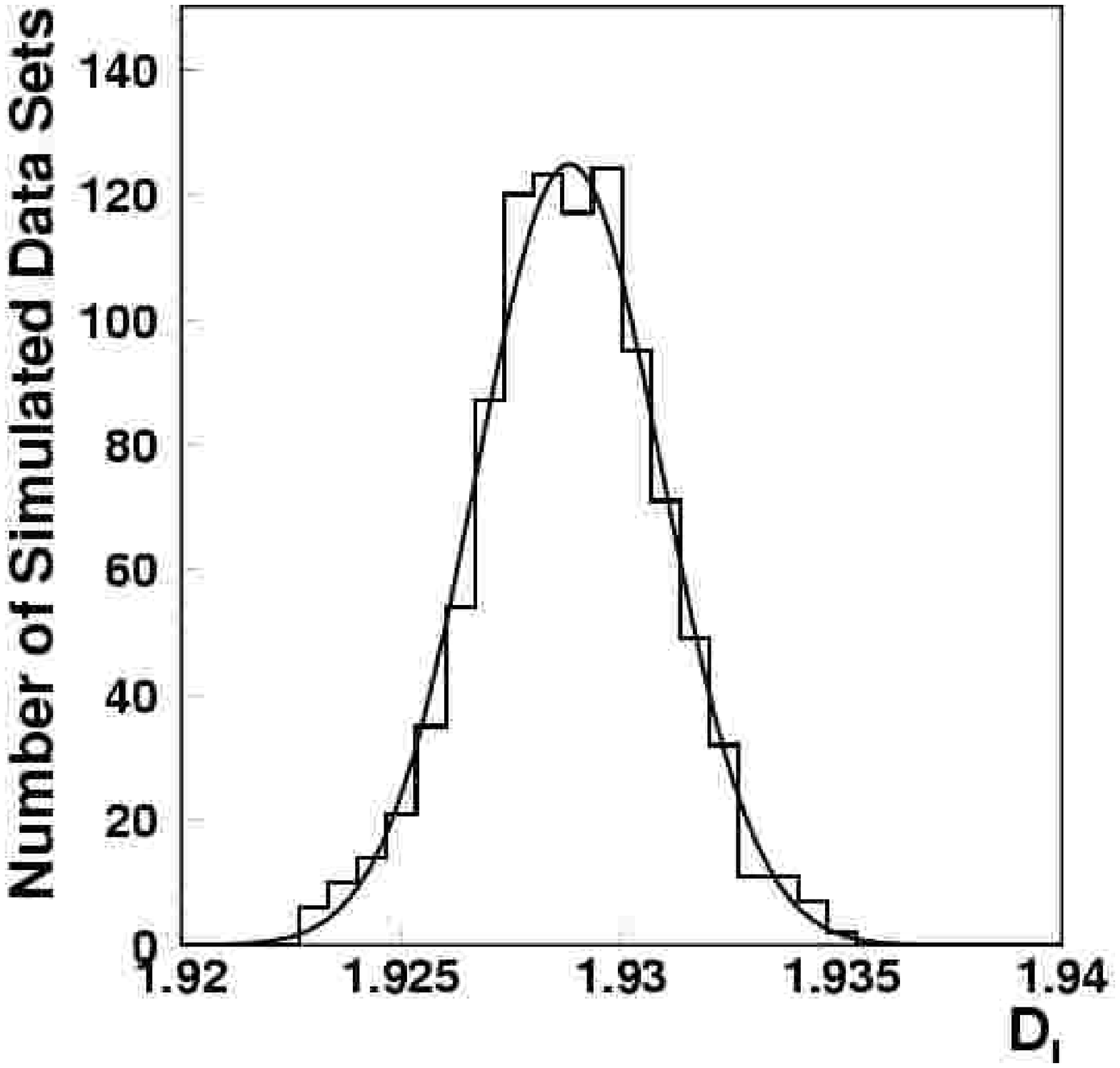}&
(b)\includegraphics[width=6.15cm]{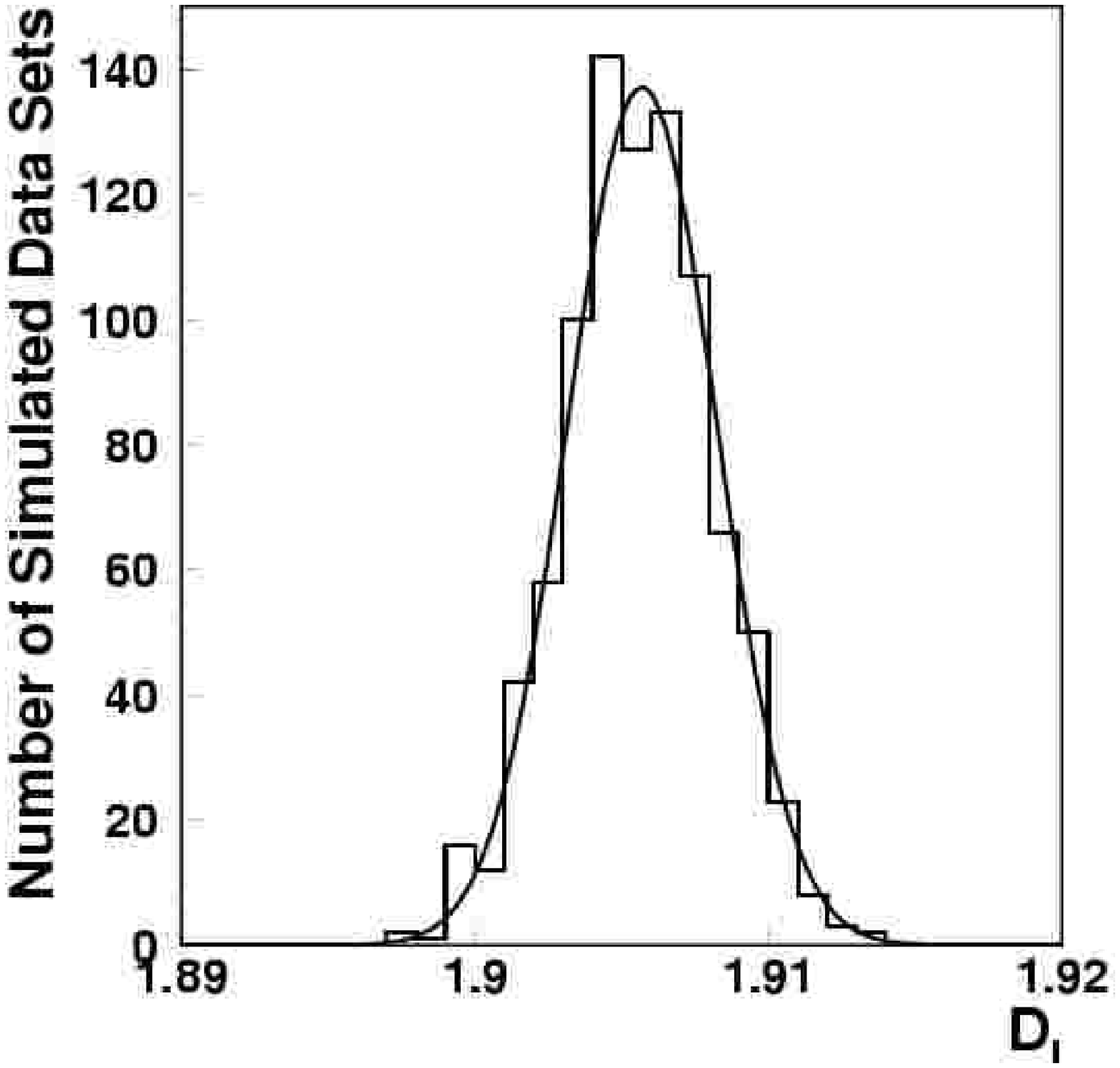}\\
\end{tabular}
\end{center}
\caption{The distribution of $D_{\rm I}$ values for simulated data sets 
with a single $\alpha$-value---(a) 
the galactic dipole source model with $\alpha=1.0$; 
(b) the galactic dipole source model with $\alpha=-1.0$.}
\label{fig:dpid}
\end{figure}

While the measurement of $D_{\rm I}$ is not necessarily the most 
sensitive tool available, it allows one to rule out any number of potential 
anisotropic source models with a single measurement.  The general scheme that
we followed is similar to what we used in the case of the dipole function. 
\begin{enumerate}
\item We calculated the value of $D_{\rm I}$ for the real data sample.  
\item We created a total of 20,000 simulated data samples, 1000 each for 0.1 
increments of $\alpha$ from -1.0 to 1.0.  In figure~\ref{fig:dpid} we can 
see that distribution of $D_{\rm I}$ values for each $\alpha$-value is
Gaussian.  
\item We constructed a curve consisting of the mean and standard
deviation of $D_{\rm I}$ for each value of $\alpha$.  
\item We then ascertained the preferred value of $\alpha$ and the 90\% 
confidence interval for each dipole source model by referring the 
intersections of the 90\% confidence interval curves with the actual value
of $D_{\rm I}$ for the real data.
\end{enumerate}

\begin{figure}[t,b,p]
\begin{center}
\begin{tabular}{c@{\hspace{0.0cm}}c}
(a)\includegraphics[width=6.15cm,height=5.85cm]{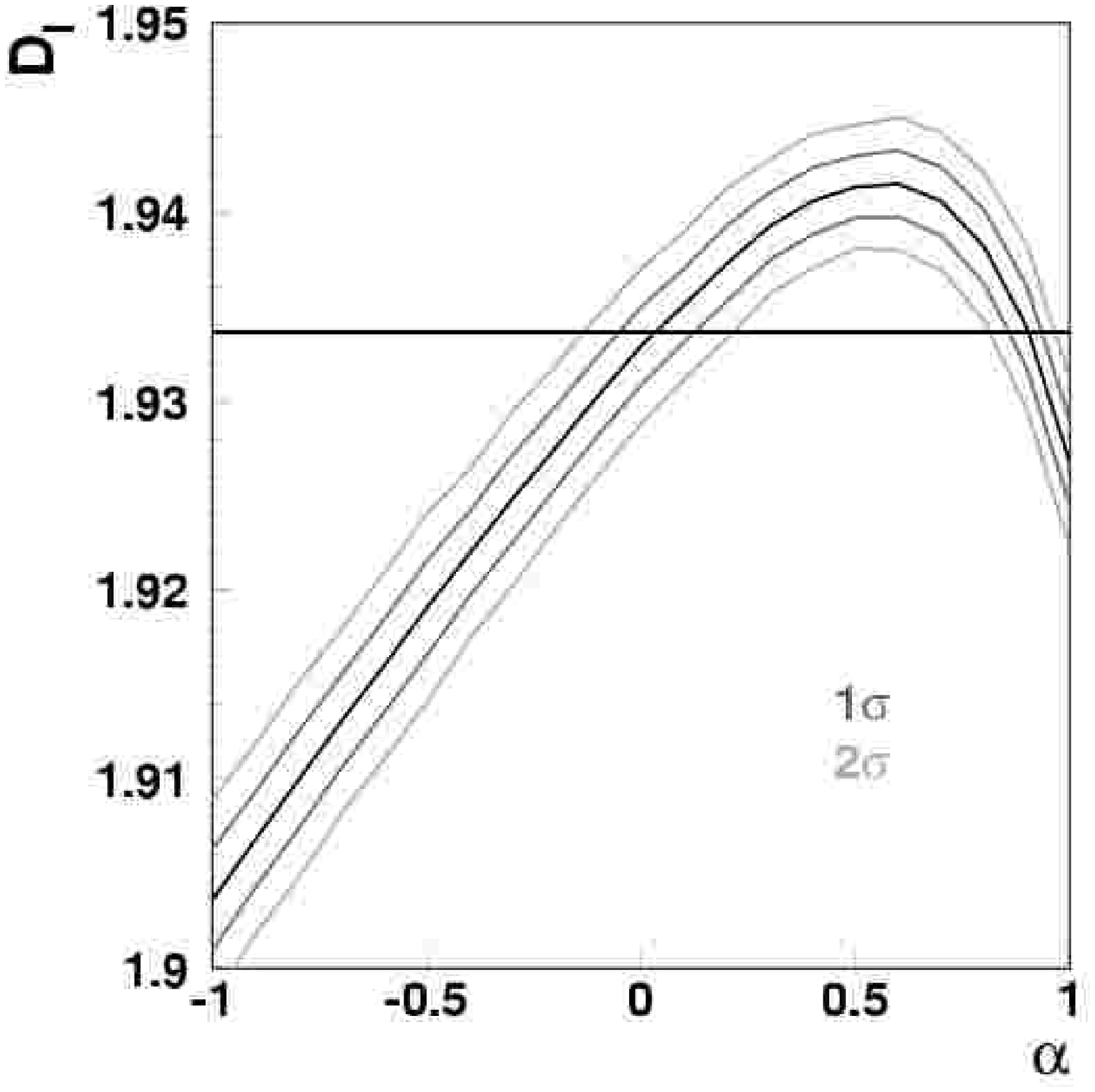}&
(b)\includegraphics[width=6.15cm,height=5.85cm]{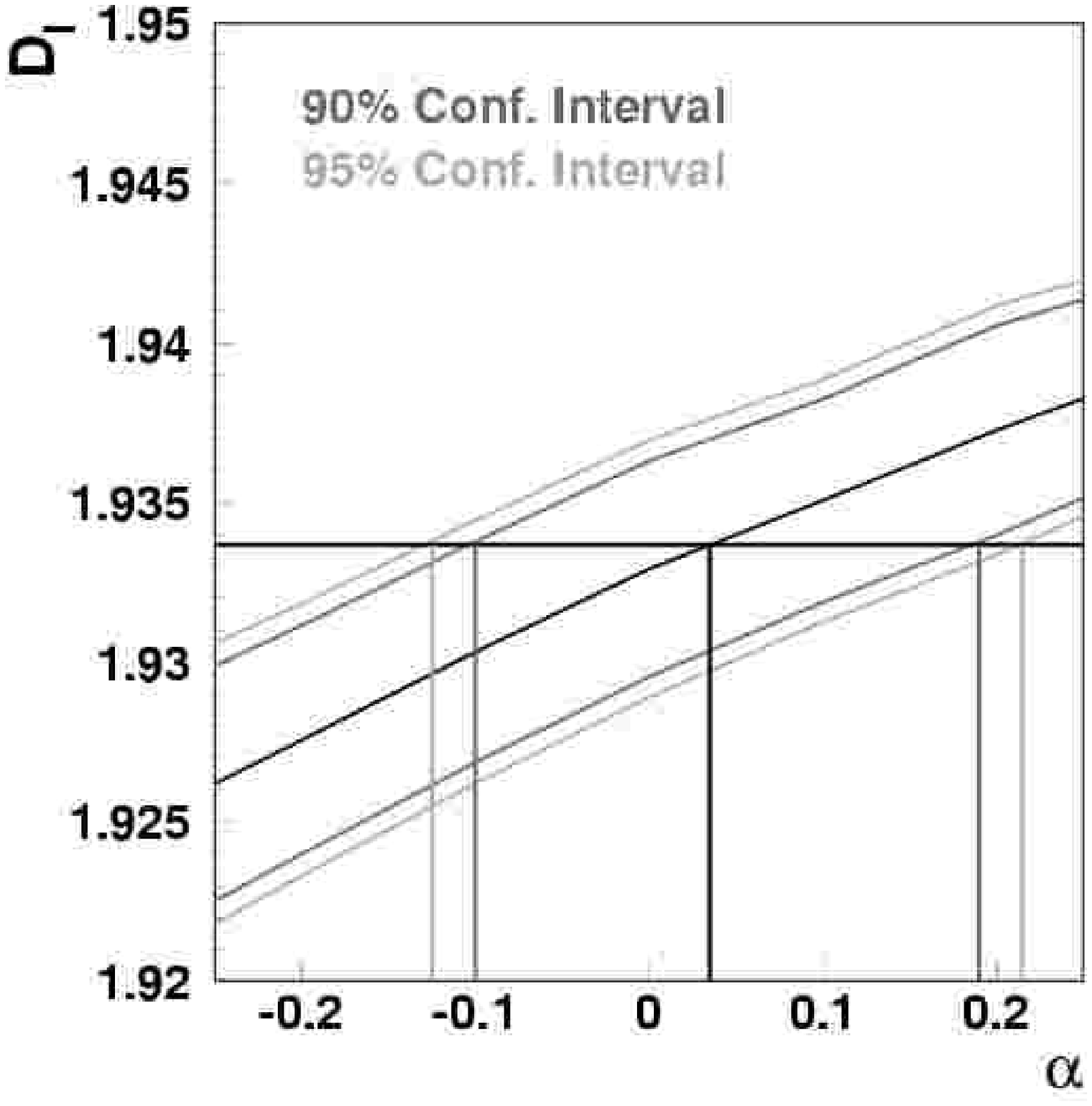}\\
(c)\includegraphics[width=6.15cm,height=5.85cm]{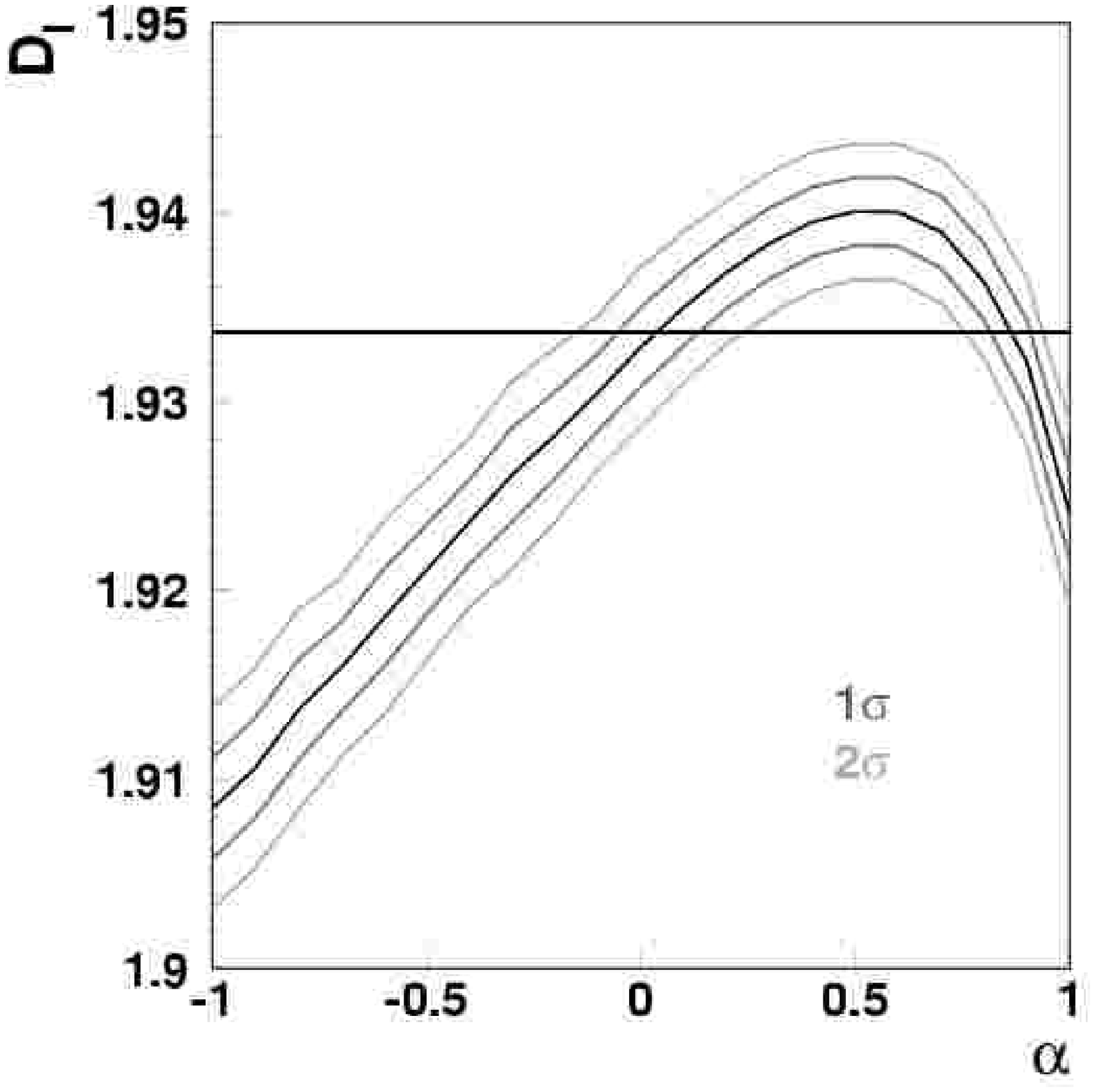}&
(d)\includegraphics[width=6.15cm,height=5.85cm]{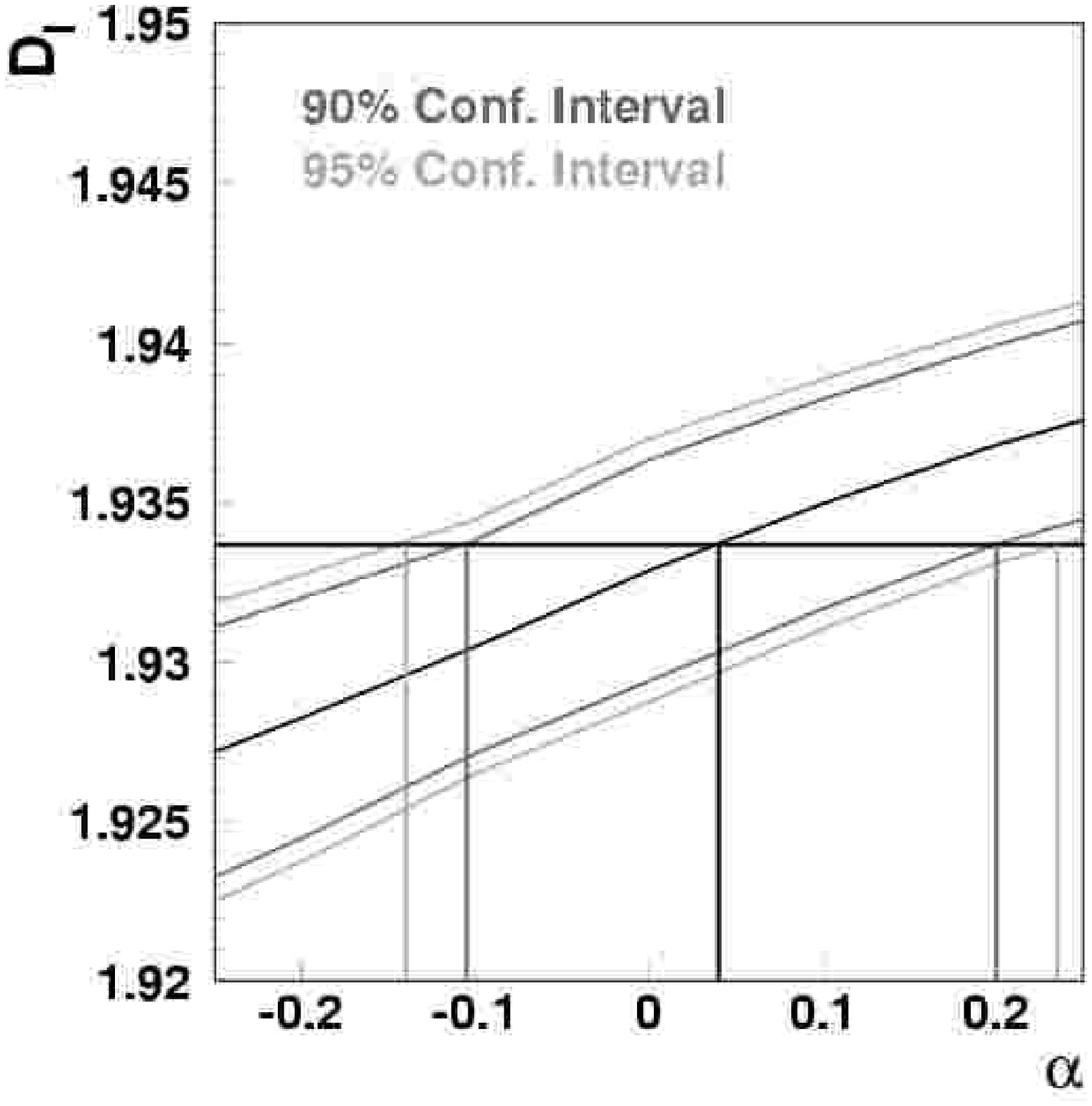}\\
(e)\includegraphics[width=6.15cm,height=5.85cm]{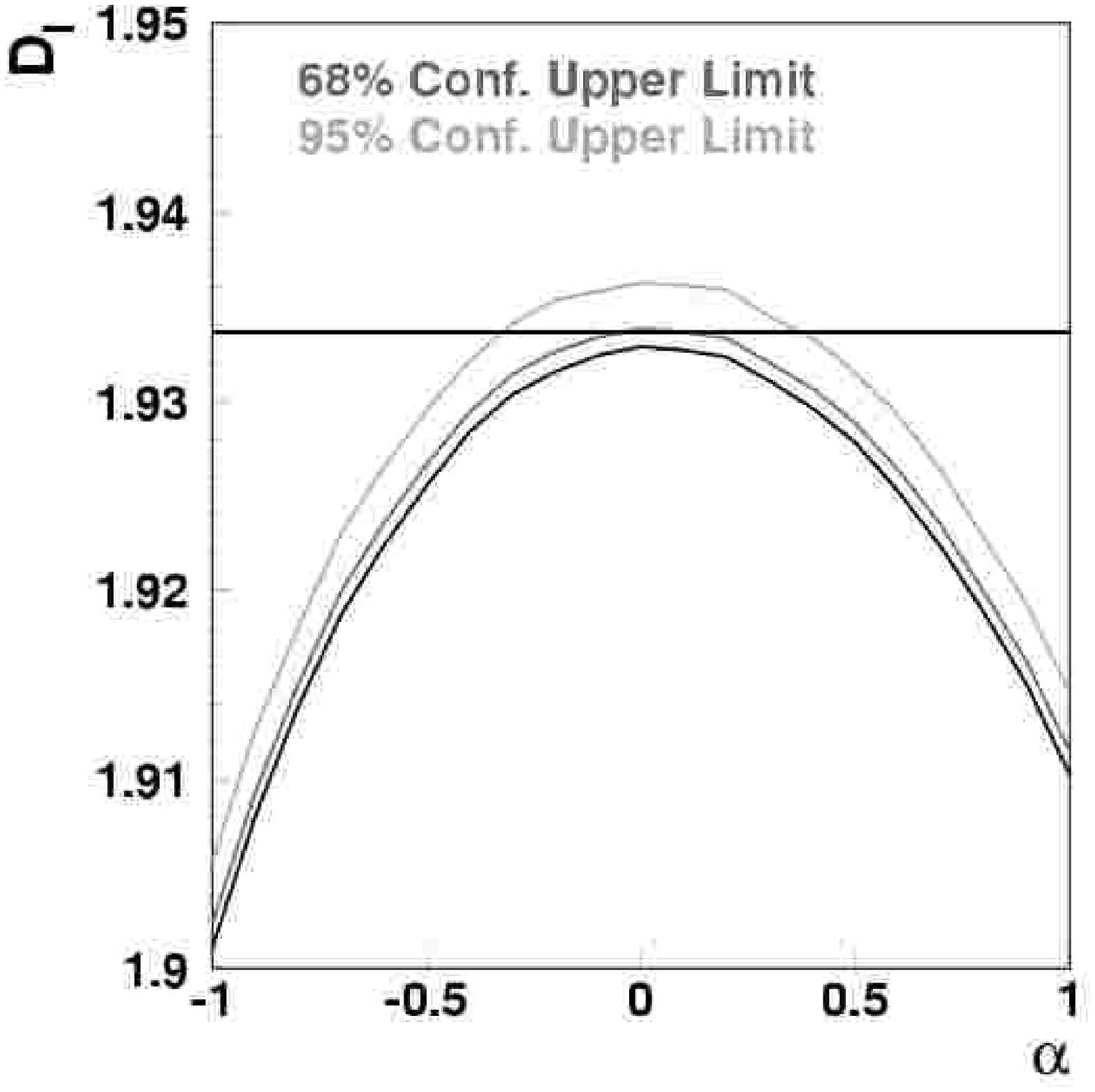}&
(f)\includegraphics[width=6.15cm,height=5.85cm]{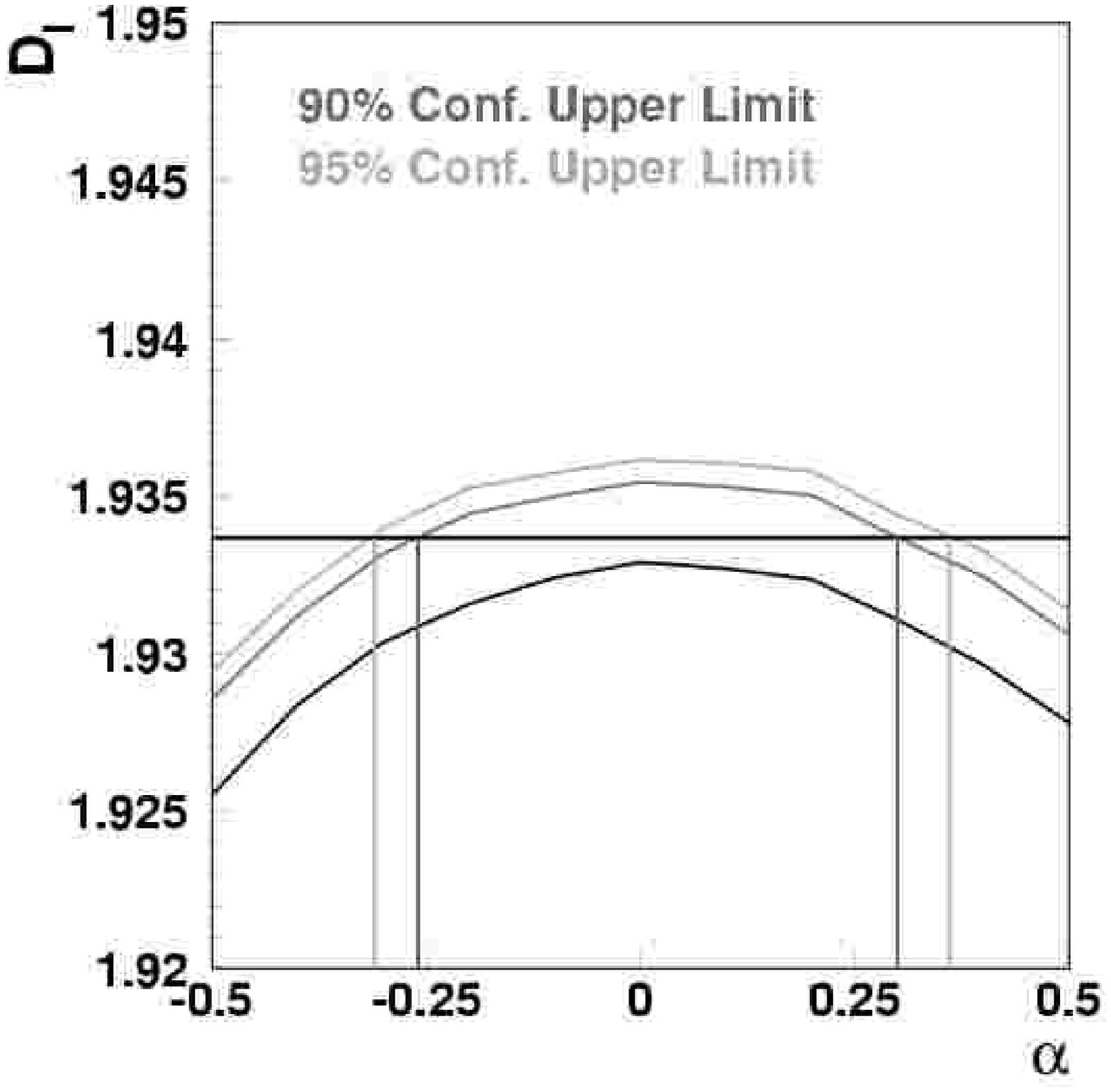}\\
\end{tabular}
\end{center}
\caption{Estimations of the value of $\alpha$ for three different dipole
source models.  The curves demonstrate the dependence of $D_{\rm I}$ upon 
$\alpha$.  The horizontal lines represent the value of $D_{\rm I}$ for
the real data---(a) 
the galactic dipole source model for $\alpha=[-1.0,1.0]$;
(b) the critical region for the galactic dipole model: $\alpha=0.035\pm0.090$ 
with a 90\% confidence interval of: $[-0.100,0.190]$;
(c) the Centaurus A dipole source model for $\alpha=[-1.0,1.0]$;
(d) the critical region for the Centaurus A dipole model:
$\alpha=0.040\pm0.095$ with a 90\% confidence interval of: $[-0.105,0.200]$; 
(e) the M87 dipole source model for $\alpha=[-1.0,1.0]$;
(f) the critical region for the M87 dipole model: $\alpha=0.020\pm0.10$ with
a 90\% confidence interval of: $[-0.26,0.30]$. 
}
\label{fig:dpi}
\end{figure}

The results for all three dipole source models are shown in 
figure~\ref{fig:dpi}.

The determination of $\alpha$ for both methods are compared in 
Table~\ref{table:comp}.  The 90\% confidence intervals for the determination
$\alpha$ via the use of $D_{\rm I}$ are substantially larger.  This is to be 
expected because the value of $D_{\rm I}$ is a single number that contains no
{\it a priori} preference for a specific source model.  Furthermore, in two
cases there is a second solution to $\alpha$ that is excluded by considering
the results of the $<\!\!\cos\theta\!\!>$ method.
The important observation is
that the results of the two methods are consistent.  
One advantage of the $D_{\rm I}$ method is that
we can state all three 90\% confidence intervals jointly, since they are all
considering only a single measurement on the real data.  In the case of the
$<\!\!\cos\theta\!\!>$ method, we would have to consider a broader confidence
interval for each individual model in order to have a simultaneous 90\% 
confidence level for all three models.
\begin{table}[t,b,p]
\begin{tabular}{|c|c|c|c|} \hline
& {\bf 1} & {\bf 2} & {\bf 3}  \\ \hline
& $\alpha$ determined  & $\alpha$ determined  & 
$\alpha$ determined  \\
{\bf SOURCE MODEL} & without considering & by the value of 
& by the value of \\ 
& angular resolution & $<\!\!\cos\theta\!\!>$ & 
$D_{\rm I}$ \\ \hline\hline
Galactic & $-0.010\pm0.055$  & $0.005\pm0.055$  & 
$0.035\pm0.09$ \\ \hline
Centaurus A & $-0.035\pm0.060$ & $-0.005\pm0.065$ & 
$0.040\pm0.095$ \\ \hline
M87 & $-0.005\pm0.045$ & $-0.010\pm0.045$ & $0.020\pm0.100$ \\ \hline
\end{tabular}
\caption{Comparison of the estimation of $\alpha$ via direct fit, 
the value of $<\!\!\cos\theta\!\!>$ for the dipole function, 
and the value of $D_{\rm I}$.}
\label{table:comp}
\end{table}
 
\section{Conclusion}

We are able to place upper limits on the value of $|\alpha|$ 
for each of our three proposed dipole source models.  However, these limits 
are not small enough to exclude the theoretical predictions 
\cite{Biermann:fd,Farrar:2000nw,Anchordoqui:2001nt}.  Also,
they do not exclude the findings of the AGASA collaboration in terms of the
intensity of the dipole effect that they observed or in terms of the energy
considered because the events in the 
dipole effect observed by the AGASA detector
possessed energies below $10^{18.5}$~eV \cite{Hayashida:1999ab}.  
Since it appears that angular resolution has little impact on the measurement
of $\alpha$ and we do not appear to be systematically limited,  
we conclude that the driving factor in making a better 
determination of $\alpha$ will simply be larger event samples.  HiRes-I
mono will continue to have the largest cumulative aperture of any single 
detector for the next three to five years, thus
it will continue to serve as an ever
more powerful tool for constraining dipole source models.

\section{Acknowledgments}
This work is supported by US NSF grants PHY 9322298, PHY 9321949, 
PHY 9974537, PHY 0071069, PHY 0098826, PHY 0140688, PHY 0245428,
by the DOE grant FG03-92ER40732,
and by the Australian Research Council. We gratefully
acknowledge the contributions from the technical staffs of our home
institutions. We gratefully acknowledge the contributions from the University
of Utah Center for High Performance Computing. The cooperation of 
Colonels E. Fisher and G. Harter, the US Army and the Dugway Proving Ground 
staff is appreciated.

\end{document}